\documentclass[%
aps,
reprint,
superscriptaddress,
 amsmath,amssymb,
pra,
]{revtex4-1}

\usepackage{textcomp}
\usepackage[utf8]{inputenc}
\usepackage{gensymb}
\usepackage{graphicx}
\usepackage{dcolumn}
\usepackage{bm}
\usepackage{mathtools}
\usepackage{hyperref}
\usepackage{color}

\usepackage[caption=false,lofdepth,lotdepth]{subfig}

\newcommand{\mb}[1]{\mathbf{#1}}
\newcommand{\ket}[1]{|#1\rangle}
\newcommand{\bra}[1]{\langle #1|}
\newcommand{\braket}[2]{\left\langle#1 |#2 \right\rangle}
\newcommand{\element}[3]{\left\langle #1\left| #2\right| #3\right\rangle} 

\begin{document}
\title{
Scattering length and effective range of microscopic two-body potentials
}
\author{Mathias Macêdo-Lima}
\affiliation{Instituto de F\'isica de S\~ao Carlos, Universidade de S\~ao Paulo, CP 369, 
13560-970 S\~ao Carlos, S\~ao Paulo, Brazil}

\author{Lucas Madeira}
\email{madeira@ifsc.usp.br}
\affiliation{Instituto de F\'isica de S\~ao Carlos, Universidade de S\~ao Paulo, CP 369, 
13560-970 S\~ao Carlos, S\~ao Paulo, Brazil}

\date{\today}

\begin{abstract}

Scattering processes are a fundamental way of experimentally probing distributions and properties of systems in several areas of physics. Considering two-body scattering at low energies, when the de Broglie wavelength is larger than the range of the potential, partial waves with high angular momentum are typically unimportant. The dominant contribution comes from $l=0$ partial waves, commonly known as $s$-wave scattering. This situation is very relevant in atomic physics, e.g. cold atomic gases, and nuclear physics, e.g. nuclear structure and matter. This manuscript is intended as a pedagogical introduction to the topic while covering a numerical approach to compute the desired quantities. We introduce low-energy scattering with particular attention to the concepts of scattering length and effective range. These two quantities appear in the effective-range approximation, which universally describes low-energy processes. We outline a numerical procedure for calculating the scattering length and effective range of spherically symmetric two-body potentials. As examples, we apply the method to the spherical well, modified P\"oschl-Teller, Gaussian, and Lennard-Jones potentials. We hope to provide the tools so students can implement similar calculations and extend them to other potentials.
\end{abstract}

\maketitle

\section{Introduction}

Scattering is commonly described as a process in which the observed object interacts with a scattering center. The objects are typically classical particles, waves or quantum particles. As to the scatterer, we generally deal with another entity, represented as a scattering potential. From a classical point of view, a collision is the most common example, where physical parameters such as the cross-section and the scattering angle arise~\cite{Griffiths2018,Newton2013}. From a quantum point of view, scattering is related to an incident wave that ``collides'' with a scattering potential, resulting in a scattered wave function.

In quantum scattering theory~\cite{Griffiths2018,Sakurai2014}, if the de Broglie wavelength is comparable to (or larger than) the range of the scattering potential, we are at the low-energy limit, in which interesting behavior appears. The properties of low-energy scattering can be described universally by two parameters: the scattering length and the effective range~\cite{Bethe1949}. Physically, the scattering length can be understood as an ``effective size'' of the target potential~\cite{Sakurai2014, Rodberg1967}. As the name suggests, the effective range may be defined as the ``real'' range of the scattering potential~\cite{Madsen2002}.  

Low-energy quantum scattering theory is important in many areas, such as nuclear, atomic, and condensed matter physics. In most physical systems, the interaction, and consequently the scattering length and effective range, is fixed. Hence, the task becomes constructing short-range potentials that describe the properties of the system. However, cold atomic gases provide a richer scenario from the point of view of varying the scattering length. The interatomic interactions between two atoms can be tuned in the laboratory via Feshbach resonances~\cite{Moerdijk1995,Randeria2014}, which allows us to explore how physical properties change as the interaction is varied. The universal behavior in the low-energy limit enables us to draw a parallel between cold atomic gases and nuclear systems, for example~\cite{Gezerlis2008,Strinati2018,Madeira2020d}.

Besides textbooks, several references aim to introduce aspects of scattering theory pedagogically. Detailed studies regarding analytical solutions in simple one-dimensional (1D) potentials are discussed in Refs.~\cite{Ribeiro2004,deCastro2011,Rizzi2016}. For 1D potentials that do not support bound states, quantities such as the reflection and transmission coefficients are typically calculated. In the tridimensional (3D) case, the quantum theory of scattering is often needed, where quantities such as the $S$-matrix, $T$-matrix, and Green's functions receive attention. Concerning analytical solutions, Ref.~\cite{Landim2022} introduces electron-positron scattering by using a covariant form of the $S$-matrix. Reference~\cite{Cavalcanti1999} gives an analytical treatment to Green's functions of delta potentials with a parallel with the scattering amplitude. Scattering length and effective range are discussed for an arbitrary angular momentum in Ref.~\cite{Pera2023}, which also presents analytical calculations for the hard-sphere, soft-sphere, spherical well and a combination of the last two potentials. As to numerical solutions, Ref.~\cite{deCarvalho2019} provides a numerical treatment to the 1D well potential by using Numerov's method, while Ref.~\cite{Viterbo2014} uses the Runge-Kutta integration method to study the variable phase equation in 3D scattering.

This work aims to construct microscopic two-body potentials to describe interactions with a desired scattering length and effective range. Since we are tuning the potential to reproduce two values, the functional form of the potential must have at least two parameters. Typically, one describes the strength of the attraction, and another is responsible for the range.

Although there are known analytical results for low-energy scattering by some potentials, it is impossible to find closed expressions for arbitrary potentials. Hence, we employed a numerical approach. This work shows how to compute the scattering length and the effective range of spherically symmetric two-body potentials by numerically solving the Schr\"odinger equation. Using Numerov's method~\cite{Caruso2014}, we integrate the equation, and the scattering theory gives the boundary conditions. Once we have the wave function, calculating the scattering length and effective range is straightforward.

We devised the structure of this work to encompass readers with several distinct goals. The main text is aimed at readers who wish to cover how the scattering length and the effective range arise in low-energy scattering theory and how to compute them numerically, but without going into the formal scattering theory. For those who want to cover scattering in more detail, in Appendix~\ref{sec:scattering_app}, we show how to obtain the Lippmann-Schwinger equation, employ Green's functions, and contour integrations in the complex plane to get the asymptotic behavior of the wave function. Covering the scattering length in graduate and undergraduate quantum mechanics courses is common, but this is not true for the effective range. Readers familiar with the scattering length may start with Sec.~\ref{sec:effective_range}, where we introduce the effective range and the shape-independent approximation. However, if the reader is proficient in low-energy scattering and seeks a numerical approach to calculate the scattering length and effective range of a given potential, Sec.~\ref{sec:numerical-procedure} onwards provides the necessary steps with relevant examples.

This work is structured as follows. In Sec.~\ref{sec:scattering_theory}, we briefly review quantum scattering theory fundamentals focusing on the low-energy limit. To keep this section concise, we start with a qualitative discussion of the asymptotic behavior of a plane wave scattered by a potential. For the readers interested in the details of scattering theory and the derivation of the equations of Sec.~\ref{sec:scattering_theory}, we provide Appendix~\ref{sec:scattering_app}. Section~\ref{sec:numerical-procedure} contains the numerical procedure to calculate the scattering length and the effective range from the zero-energy solutions of the Schr\"odinger equation. In Sec.~\ref{sec:examples}, we apply the method to the spherical well, modified P\"oschl-Teller (mPT), Gaussian, and Lennard-Jones (LJ) potentials as examples to illustrate the method. Finally, in Sec.~\ref{sec:conclusion}, we present our final considerations.

\section{Scattering theory}\label{sec:scattering_theory}

We briefly present some quantum scattering theory results to define the scattering length, following textbook references~\cite{Griffiths2018, Sakurai2014}. We are also interested in the effective-range expansion, which is not typically covered in undergraduate quantum mechanics courses. For this reason, we employ arguments from references~\cite{Rodberg1967,Madsen2002}. The reader is referred to the references mentioned above for a detailed derivation of the results.

\begin{figure}[!htb]
\centering
\includegraphics[width=\linewidth]{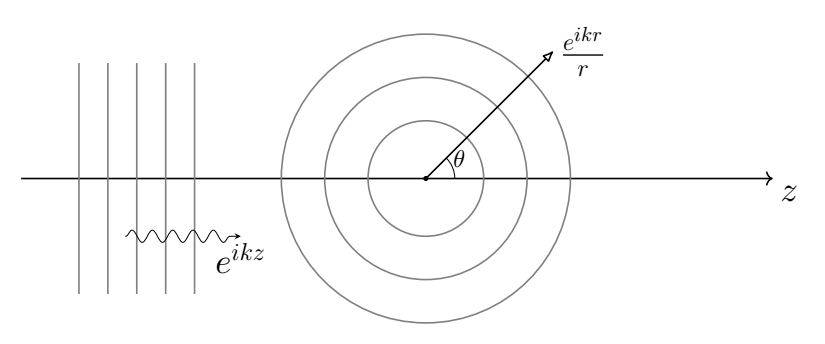}
\caption{Scheme of a scattering process. We chose our coordinate system such that the incoming particle has its momentum in the $z$-direction, $\mb{k}=k\hat{\mb{z}}$. A plane wave $e^{i\mb{k}\cdot{\mb{r}}}=e^{ikz}$ travels in space until it encounters a scattering potential. It produces an outgoing spherical wave $e^{ikr}/r$ at large distances. The scattering angle $\theta$ defines the direction in which the momentum was transferred after the scattering process.}
\label{fig:scat_drawing} 
\end{figure}

Formally, a scattering process is described as a transition from one quantum state to another. We are interested in a particle, initially far away from the region where it will be scattered, moving toward the scattering center. The initial state $\ket{i}$ is assumed to be a plane wave $\ket{\mb{k}}$ since it is a free particle. The final state $\ket{f}$ is the result of the action of a scattering potential on $\ket{i}$ which, at large distances, is an outgoing spherical wave as illustrated in Fig.~\ref{fig:scat_drawing}. The initial state $\ket{i}$ satisfies the eigenvalue equation for the free-particle Hamiltonian $H_0 = \mb{p}^2/2m$,
\begin{equation}\label{eq:free_eigenvalue_eq}
H_0\ket{i} = E_i\ket{i} = \frac{\hbar^2\mb{k}^2}{2m}\ket{\mb{k}}.
\end{equation}

We choose our coordinate system such that the scattering center is at the origin, and the position vector is given by $\mb{r} = x\hat{\mb{x}}+y\hat{\mb{y}}+z\hat{\mb{z}}$. Scattering is taken into account by introducing a potential $V(\mb{r})$ that acts on a finite range $r<R$, where $r\equiv |\mb{r}|$. The Hamiltonian in this region is
\begin{equation}
H = H_0 + V(\mb{r}).
\end{equation}
Our goal is to understand the action of $H$ on free-particle states and how the scattering occurs. In other words, we want to know how the initial state $\ket{i}$ transitions to a final state $\ket{f}$.

The initial state is given by a plane wave, $\ket{i} = \ket{\mb{k}}$. We consider the particle to be inside a cubic box of side $L$, such that the free particle is represented in the position basis as
\begin{equation}
\braket{\mb{r}}{\mb{k}} = \mathcal{N}\,e^{i\mb{k}\cdot\mb{r}} = \frac{e^{i\mb{k}\cdot\mb{r}}}{L^{3/2}}.
\end{equation}
The factor $\mathcal{N} = L^{-3/2}$ guarantees the normalization of the plane wave, and the dot product $\mb{k}\cdot\mb{r}$ defines the scalar product between the momentum $\mb{k} = \mb{p}/\hbar = k_x\hat{\mb{x}}+k_y\hat{\mb{y}}+k_z\hat{\mb{z}}$ and the position. At the end of our calculations, we must take $L\to \infty$ to guarantee the continuum character of the state.

Although $V$ is time-independent, we may tackle our problem with a time-dependent formalism by considering the potential as an ``adiabatic switch''. The potential is  ``turned on'' while $r<R$, inside the scattering region, and rapidly ``turned off'' when $r>R$. Furthermore, the scattering may be assumed to be elastic. The transition is from $\ket{i}$ to a group of continuum states $\ket{f}$ with energies $E_f$, where both $\ket{i}$ and $\ket{f}$ are eigenstates of $H_0$.

The procedure outlined in the last paragraph is carried out in detail in Appendix~\ref{sec:scattering_app}. Here, we reproduce the result for the asymptotic behavior of the wave function,
\begin{equation}\label{eq:scat-wavefunction}
\psi(\mb{r}) \xrightarrow{\text{large }r} \frac{1}{L^{3/2}}\left[e^{i\mathbf{k}\cdot \mathbf{r}}+\frac{e^{ikr}}{r}f(\mb{k}',\mb{k})\right].
\end{equation}
This is a quantitative description of what we saw in Fig.~\ref{fig:scat_drawing}: at large distances, the wave function combines the incident plane wave and a scattered spherical wave. Notice that the factor $f(\mb{k}',\mb{k})$ is multiplying the spherical wave in Eq.~(\ref{eq:scat-wavefunction}). This term is called scattering amplitude, and it indicates how much of the incident wave was scattered. It depends on the initial $\mb{k}$ and final $\mb{k}'$ momenta,
\begin{equation}
f(\mb{k}',\mb{k}) = -\frac{mL^3}{2\pi\hbar^2}\int d^3r' \braket{\mb{k}'}{\mb{r}'}V(\mb{r}')\braket{\mb{r}'}{\psi}.
\end{equation}
Equation~(\ref{eq:scat-wavefunction}) contains a large amount of information. Without solving the Schrödinger equation explicitly, we arrived at a result for the wave function at large distances with only a few assumptions about the scattering potential.

\subsection{Partial waves expansion}

To obtain Eq.~(\ref{eq:scat-wavefunction}), we assumed the scattering potential $V(\mb{r})$ to be real, finite-ranged, and local. Now, we consider one last restriction: that it is spherically symmetric, i.e., $V$ is invariant under rotations. Hence, we can write $V(\mb{r}) = V(r)$. We must then solve Schrödinger's equation for a potential $V(r)$ such that $V(0<r<R) \ne 0$ and $V(r>R) = 0$. In the scattering region ($0<r<R$), we write 
\begin{equation}
-\frac{\hbar^2}{2m}\nabla^2\psi + V(r)\psi = E\psi.
\end{equation}
The total energy is $E = \hbar^2k^2/2m$. A straightforward way of seeing this is to notice that this is the energy of the particle in the region $r>R$, where $V(r)=0$. Since the scattering process is elastic, the total energy is conserved.

The wave function $\psi = \psi_\mb{k}(r,\theta)$ depends only on the momentum $\mb{k}$, the position $r$ and the scattering angle $\theta$, due to the spherical symmetry. We want the solution to obey the asymptotic behavior of Eq.~(\ref{eq:scat-wavefunction}), which we now write as
\begin{equation}\label{eq:scat-wavefunction2}
\psi_{\mb{k}}(r,\theta) \xrightarrow{\text{large }r} \mathcal{N}\left[e^{ikz}+\frac{e^{ikr}}{r}f(\theta)\right],
\end{equation}
where $\mathcal{N}$ is a normalization constant. We also considered $\mb{k}\cdot\mb{r} = kz$, corresponding to an incident plane wave in the $z$ direction.

Due to the spherical symmetry of $V(r)$, it is convenient to employ spherical coordinates. We write the Laplace operator in spherical coordinates and $\psi = \psi(r,\theta,\phi)$, for which we derive the eigenvalue equation
\begin{equation}
\label{eq:sch_spherical}
    \left(-\frac{\hbar^2}{2m}\frac{1}{r}\frac{\partial^2}{\partial r^2}r + \frac{L^2}{2mr^2} +V(r)\right) \psi(r,\theta,\phi) = E\psi(r,\theta,\phi).
\end{equation}
The angular momentum operator $L$ is such that
\begin{equation}
L^2 = -\hbar^2\left(\frac{1}{\sin\theta} \frac{\partial}{\partial \theta}\sin\theta \frac{\partial}{\partial \theta} + \frac{1}{\sin^2\theta} \frac{\partial^2}{\partial \phi^2}\right).
\end{equation}
Its $z$-component is given by
\begin{equation}
L_z=-i\hbar\frac{\partial}{\partial \phi}.
\end{equation}
We then construct a complete set of eigenfunctions related to the commuting observables $H, L^2$, and $L_z$ with eigenvalues $E$, $l(l+1)\hbar^2$, and $m\hbar$, respectively,
\begin{eqnarray}
H \psi(r,\theta,\phi) &=& E \psi(r,\theta,\phi), \nonumber\\
L^2 \psi(r,\theta,\phi) &=& l (l+1)\hbar^2  \psi(r,\theta,\phi), \nonumber\\
L_z \psi(r,\theta,\phi) &=& m\hbar \psi(r,\theta,\phi).
\end{eqnarray}

We propose a separable solution of the form
\begin{equation}
\label{eq:psi_sep}
\psi(r,\theta,\phi) = A_l(r)Y_l^m(\theta,\phi),
\end{equation}
where the $A_l(r)$ are radial functions and the $Y_l^m(\theta,\phi)$ are the spherical harmonics. The angular part of the wave function, which depends on $\theta$ and $\phi$, for any spherically symmetric potential corresponds to the spherical harmonics. Typically, students first encounter this property while studying the particular case of the hydrogen atom. This is a consequence of the potential commuting with $L^2$ and $L_z$, $[V,L^2]=0$ and $[V,L_z]=0$, which is true for any potential that depends only on the radial coordinate $r$.

We perform a change of variables $A_l(r)=u_l(r)/r$, which is convenient given that there is a first derivative in the radial equation for $A_l(r)$, and this change removes it from the corresponding equation for $u_l(r)$. Substituting Eq.~(\ref{eq:psi_sep}) into (\ref{eq:sch_spherical}) yields an equation that depends only on the radial coordinate,
\begin{equation}\label{eq:radial_equation}
    \left(\frac{d^2}{dr^2} + k^2 - U(r) - \frac{l(l+1)}{r^2}\right)u_l(r) = 0,
\end{equation}
where $U(r) = 2mV(r)/\hbar^2$. At the origin, we require that $A_l(r)$ is finite. Since $A_l(r)=u_l(r)/r$, we need $u_l(0) =0$.

We have a free particle in the region $r>R$ since $U(r) = 0$. In this case, the solutions of Eq.~(\ref{eq:radial_equation}) can be written in terms of the spherical Bessel functions of the first and second kind, usually denoted by $j_l(x)$ and $n_l(x)$, respectively. The solution is of the form
\begin{equation}
\label{eq:sph_bessel}
u_l(r) = c_l'rj_l(kr) + c_l''rn_l(kr), 
\end{equation}
where $c_l'$ and $c_l''$ are constants.
It is convenient to write the solution in terms of a linear combination of spherical Bessel functions,
\begin{eqnarray}
\label{eq:hankel}
h_l^{(1)}(x)  &=& j_l(x) + in_l(x),\nonumber\\
h_l^{(2)}(x)  &=& j_l(x) - in_l(x),
\end{eqnarray}
where $h_l^{(1)}$ and $h_l^{(2)}$ are called spherical Hankel functions of the first and second kind, respectively.
Thus, Eq.~(\ref{eq:sph_bessel}) can be written as
\begin{equation}\label{eq:outside_solution}
u_l(r)  = c_l^{(1)}rh_l^{(1)}(kr) + c_l^{(2)}rh_l^{(2)}(kr),
\end{equation}
where $ c_l^{(1)}$ and $ c_l^{(2)}$ are constants.

We know that the solution for a free particle, a plane wave $e^{ikz}=e^{ikr\cos\theta}$, contains components with all possible values of the angular momentum, $l=0,1,2,...$, and we wish to decompose it into a sum of different angular momentum components. This can be done with Rayleigh's formula~\cite{Arfken2005},
\begin{equation}
\label{eq:rayleigh}
e^{ikr\cos\theta} = \sum_{l=0}^\infty i^l(2l+1)j_l(kr)P_l(\cos\theta),
\end{equation}
where the $P_l$ are Legendre polynomials. Notice that the plane wave, and consequently its expansion, does not depend on $\phi$. This is due to the spherical symmetry of the problem. If we think in terms of spherical harmonics, their $\phi$ dependence is $Y_l^m \propto e^{im\phi}$. The absence of $\phi$ indicates that we must set $m=0$, which leads directly to the Legendre polynomials since $Y_l^0(\theta,\phi) \propto P_l(\cos\theta)$.

Let us analyze the asymptotic behavior of Eq.(\ref{eq:rayleigh}), when $r\to \infty$. Equation~\ref{eq:hankel} allows us to write
\begin{equation}
\label{eq:bessel_hankel}
j_l(x)=\frac{h_l^{(1)}(x)+h_l^{(2)}(x)}{2}.
\end{equation}
The spherical Hankel functions obey the following limits~\cite{Arfken2005}
\begin{eqnarray}
\label{eq:hankel_asym}
h_l^{(1)}(x)&\xrightarrow{\text{large }x}& (-i)^{l+1}\frac{e^{ix}}{x},\nonumber\\
h_l^{(2)}(x)&\xrightarrow{\text{large }x}& i^{l+1}\frac{e^{-ix}}{x}.
\end{eqnarray}
Gathering all this information, we have that
\begin{equation}
\label{eq:rayleigh_asym}
e^{ikr\cos\theta} \xrightarrow{\text{large }r} \sum_{l=0}^\infty \frac{(2l+1)}{2ikr}\left[ 
e^{ikr}-(-1)^l e^{-ikr}
\right]P_l(\cos\theta).
\end{equation}
The first term inside the square brackets represents an outgoing spherical wave, while the second is related to an incoming spherical wave.

In the region $r>R$ we have not only the incident plane wave, but also the spherical wave produced by the scattering center. Motivated by Eq.~(\ref{eq:rayleigh}) and its asymtoptic behavior, Eq.~(\ref{eq:rayleigh_asym}), we write the solution for every $r>R$ as
\begin{equation}
    \psi(r,\theta) = \mathcal{N}\sum_{l=0}^\infty i^l(2l+1)\frac{u_l(r)}{r}P_l(\cos\theta),
\end{equation}
where $\mathcal{N}$ is a normalization constant. Let us analyze the behavior of $\psi(r,\theta)$ when $r\to \infty$. Equations~(\ref{eq:outside_solution}) and (\ref{eq:hankel_asym}) yield
\begin{flalign}
\label{eq:asym_psi_ul}
&
\psi(r,\theta)
\xrightarrow{\text{large }r}&\nonumber\\
&
\mathcal{N}\sum_{l=0}^\infty \frac{(2l+1)}{ikr} 
\left[c_l^{(1)} e^{ikr} - (-1)^l c_l^{(2)} e^{-ikr}\right] P_l(\cos\theta).&
\end{flalign}

Let us pause to consider the implications of this result. Equation~(\ref{eq:rayleigh_asym}) describes the asymptotic behavior of the wave function for a plane wave without being scattered, while Eq.~(\ref{eq:asym_psi_ul}) does the same, but in a situation where scattering could have taken place. If we take $\mathcal{N}=1$ and $c_l^{(1)}=c_l^{(2)}=1/2$, then both equations are the same. This is not surprising since this particular choice makes the radial function the same as the one for a free particle, $u_l(r)/r=j_l(kr)$ [see Eqs.(\ref{eq:outside_solution}) and (\ref{eq:bessel_hankel})]. However, this situation arises only because the coefficients in front of the outgoing spherical wave and the incoming one, $c_l^{(1)}$ and $c_l^{(2)}$ respectively, are the same. If we take $c_l^{(1)}\neq c_l^{(2)}$, then scattering certainly took place. Moreover, the proportion of outgoing to incoming spherical waves, given by the ratio of these two coefficients, can be used to quantify the impact of the scattering potential on the free particle.

This motivates us to introduce a new quantity related to the ratio between the constants in front of $e^{ikr}/r$ and $e^{-ikr}/r$,
\begin{equation}\label{eq:scattering_ratio}
    \frac{c_l^{(1)}}{c_l^{(2)}}  = S_l(k) = e^{2i\delta_l(k)},
\end{equation}
where the $S_l(k)$ are called scattering matrix components and the $\delta_l(k)$ are called  phase shifts. Notice that $\delta_l(k)$ depends explicitly on the momentum or, equivalently, on the energy. The free particle case, $c_l^{(1)}=c_l^{(2)}=1/2$, corresponds to a phase shift $\delta_ l(k)=0$.  Writing the ratio of two constants as a complex exponential may seem strange, but the reason will be apparent shortly.

Let us express the asymptotic wave function, Eq.~(\ref{eq:asym_psi_ul}), in terms of the phase shifts, Eq.~(\ref{eq:scattering_ratio}),
\begin{flalign}
\label{eq:asymptotic_hankel_wavefunction}
&
\psi(r,\theta)
\xrightarrow{\text{large }r}&\nonumber\\
&
\mathcal{N}\sum_{l=0}^\infty \frac{(2l+1)}{ikr} c_l^{(2)}
\left[e^{2i\delta_l} e^{ikr} - (-1)^l e^{-ikr}\right] P_l(\cos\theta).&
\end{flalign}
Recall that our analysis started with the idea that a plane wave is scattered and produces a spherical wave, the latter multiplied by the scattering amplitude $f(\theta)$, as given by Eq.~(\ref{eq:scat-wavefunction2}). Now, we can connect the scattering amplitude and the phase shifts. For that, we recast Eq.~(\ref{eq:scat-wavefunction2}) using the asymptotic limit of Rayleigh's formula, Eq.~\ref{eq:rayleigh_asym},
\begin{flalign}
\label{eq:scattering_amp_rayleigh_asym}
&
\psi(r,\theta)
\xrightarrow{\text{large }r}
\mathcal{N}\left\{
\left[
\sum_{l=0}^\infty \frac{(2l+1)}{2ikr}\left(
e^{ikr}-(-1)^l e^{-ikr}
\right)\times \right.\right.&\nonumber\\
&
\left.
P_l(\cos\theta)
\Bigg]
+f(\theta)\frac{e^{ikr}}{r}
\right\}.&
\end{flalign}
Comparison between Eqs.~(\ref{eq:asymptotic_hankel_wavefunction}) and (\ref{eq:scattering_amp_rayleigh_asym}) allows us to write the scattering amplitude as a function of the phase shifts,
\begin{equation}
    f(\theta) = \sum_{l=0}^\infty (2l+1) \frac{(e^{2i\delta_l} - 1)}{2ik}P_l(\cos\theta).
\end{equation}
The factor $(e^{2i\delta_l}-1)/2ik$ is referred to as the partial wave amplitude $f_l(k)$, which may be rewritten as 
\begin{equation}
    f_l(k) = \frac{e^{2i\delta_l} - 1}{2ik}= \frac{e^{i\delta_l}\sin\delta_l}{k},
\end{equation}
or
\begin{equation}
\label{eq:fl}
    f_l(k) = \frac{e^{i\delta_l}\sin\delta_l}{k} = \frac{1}{k\cot\delta_l-ik},
\end{equation}
whichever is more convenient. Using Eq.~(\ref{eq:scattering_ratio}), we may also express the partial wave amplitude in terms of $S_l(k)$ as
\begin{equation}{\label{eq:partial_wave_amplitude-scattering_matrix_element}}
    S_l(k) = 1+2ikf_l(k).
\end{equation}

\begin{figure}[!htb] 
    \centering
    \begin{minipage}{0.5\textwidth}
        \centering
        \subfloat[Attractive potential: $\delta_0(k)>0$. \label{fig:phase_shift_a}]{\includegraphics[width=\textwidth]{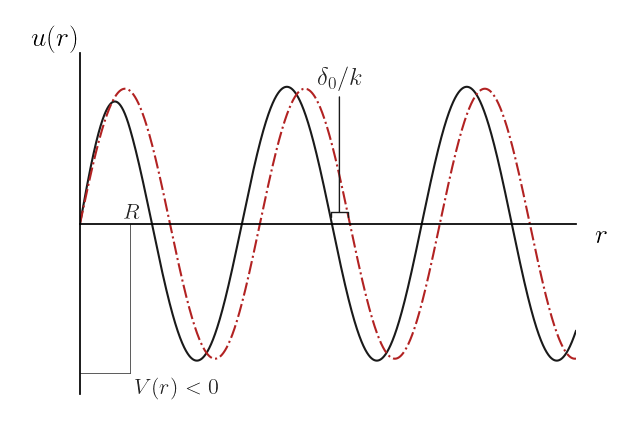}}
    \end{minipage}\hfill
    \begin{minipage}{0.5\textwidth}
        \centering
        \subfloat[Repulsive potential: $\delta_0(k)<0$. \label{fig:phase_shift_b}]{\includegraphics[width=\textwidth]{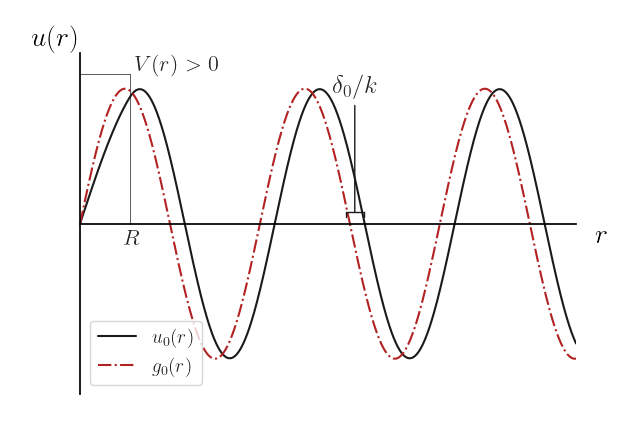}}
    \end{minipage}
    \caption{Scattered radial solution $u_0(r)$ and the free ($V=0)$ radial solution $g_0(r)$ to highlight the phase shift. At low energies, an attractive potential ($V(r)<0$) pulls the wave function, resulting in $\delta_0(k)>0$. The repulsive potential ($V(r)>0$) pushes $u_0(r)$ away, resulting in $\delta_0(k)<0$.}
    \label{fig:phase_shift}
\end{figure}

Now we can attribute physical meaning to the partial wave scattering amplitude and the phase shifts. Conservation of the probability during scattering tells us that, at large distances, the only thing that can change is the phase of the wave function (with respect to the incident wave). The difference between the phases is the phase shift $\delta_l(k)$. When there is no scattering, $V=0$, $\delta_l(k)=0$ and $f_l(k)=0$, meaning that we have the free particle solution in all space, indicated as a dot-dashed line in Figure~\ref{fig:phase_shift}. For a potential $V\neq 0$, the radial solution for $r<R$ will depend on the details of the potential. However, we have a free particle solution outside the range $R$ of the potential, $V(r>R)=0$. Hence, what happens inside the range of the potential determines the phase shift observed outside of it. An attractive potential ``pulls'' the particle toward the origin, Fig.~\ref{fig:phase_shift_a}, while a repulsive potential ``pushes'' it away, Fig.~\ref{fig:phase_shift_b}. The mathematical advantage of this formulation is that we describe the whole process in terms of a real quantity $\delta_l(k)$, and we reduce our problem to calculating it.

To compute the phase shifts, we can use two properties of wave functions: they are continuous, and so are their first derivatives. Hence, we need to match the inside ($r<R$) and outside ($r>R$) solutions and their derivatives at $r=R$. We write the radial solution in the region $r>R$ as 
\begin{equation}\label{eq:low_energy_partial_wave-amplitude}
    A_l(r) = \frac{u_l(r)}{r} = \frac{1}{2}e^{2i\delta_l}h_l^{(1)}(kr) + \frac{1}{2}h_l^{(2)}(kr) 
\end{equation}
in terms of spherical Hankel functions, and
\begin{equation}
\label{eq:Al_bessel}
    A_l(r) = e^{i\delta_l}(\cos\delta_lj_l(kr) - \sin\delta_l n_l(kr))
\end{equation}
in terms of spherical Bessel functions. To avoid calculating the normalization of the wave function, we can work with the ratio $u_l'(r)/u_l(r)$ so that it cancels out. This quantity is known as a logarithmic derivative because
\begin{equation}
\frac{d}{dx} \ln f(x)=\frac{f'(x)}{f(x)}.
\end{equation}
We define a dimensionless logarithmic derivative by including a factor of $r$. At $r=R^-$ (at the range $R$ of the potential coming from inside), it is a constant which depends on the expression of $V(r)$,
\begin{equation}
\label{eq:log_der}
    \beta_l= \left[r\frac{u'_l(r)}{u_l(r)}\right]_{r=R^-}.
\end{equation}
The phase shift can be determined by equating the logarithmic derivative at $r=R^-$ with the outside solution (at $r=R^+$) given by Eq.~(\ref{eq:Al_bessel}),
\begin{eqnarray}
    \beta_l &=& \left[r\frac{u'_l(r)}{u_l(r)}\right]_{r=R^+}\nonumber\\
    &=&1+ kR\left[\frac{\cos\delta_l j'_l(kR)-\sin\delta_l n'_l(kR)}{\cos\delta_l j_l(kR) - \sin\delta_l n_l(kR)}\right],
\end{eqnarray}
where the notation $j'_l(kR)$ means the derivative of $j_l$ with respect to $kr$ evaluated at $kR$, and similarly for $n'_l(kR)$. After some algebra, we arrive at
\begin{equation}
\cot\delta_l(k) = \frac{kR\,n_l'(kR)-(\beta_l-1)\,n_l(kR)}{kR\,j_l'(kR)-(\beta_l-1) \,j_l(kR)}.
\end{equation}
This is an analytic expression to calculate the $l$-th partial wave phase-shift $\delta_l(k)$ provided we know the inside solution to compute the constant $\beta_l$. This result is a cornerstone for the numerical procedure we will treat later in this article.

\subsection{The low-energy limit and the scattering length}

From Eq.~(\ref{eq:sch_spherical}), it is possible to see that the particle is subjected to an effective potential for the $l$-th partial wave of
\begin{eqnarray}
V_{\rm eff}(r)=V(r)+\frac{\hbar^2}{2m}\frac{l(l+1)}{r^2},
\end{eqnarray}
where the second term on the right-hand side is a repulsive centrifugal barrier. It is absent for $l=0$, becoming more repulsive as we increase the value $l$ of the considered partial wave. This work focuses on low-energy scattering, where the reduced wavelength $\lambdabar = \lambda/2\pi = 1/k$ is much larger than the potential range, that is $\lambdabar \gg R$ or $kR \ll 1$. If the energy is close to zero, $E\approx 0$, then the particle cannot overcome the centrifugal barrier. In this case, the partial waves with $l>0$ are unimportant, and the $l=0$ component is dominant in understanding low-energy scattering.

In this low-energy scenario, we consider partial waves with $l\ne 0$ to vanish, and the resulting $l=0$ term is referred to as ``$s$-wave''. Thus, the $s$-wave scattering radial component is given by Eq.~(\ref{eq:Al_bessel}),
\begin{eqnarray}\label{eq:l=0_radial_solution}
A_0(r)=\frac{u_0(r)}{r} &=& e^{i\delta_0}(\cos\delta_0 j_0(kr) - \sin\delta_0 n_0(kr))\nonumber\\
&=&e^{i\delta_0}\left[\frac{1}{kr}\sin(kr+\delta_0)\right],
\end{eqnarray}
where we used
\begin{eqnarray}
j_0(x)&=&\frac{\sin x}{x},\nonumber\\
n_0(x)&=&-\frac{\cos x}{x}.
\end{eqnarray}

Schrödinger's equation for the radial solution becomes very simple in this situation. Outside the range of the potential, $V(r>R)=0$. Moreover, there is no centrifugal barrier since $l=0$, and for low-energy scattering, $k\approx 0$. Hence, Eq.~(\ref{eq:radial_equation}) reduces to
\begin{equation}
u_0''(r) = 0.
\end{equation}
The solution can be written as
\begin{equation}
\label{eq:line}
u_0(r)=c(r-a),
\end{equation}
where $c$ and $a$ are constants. Its logarithmic derivative, given by Eq.~(\ref{eq:log_der}), is
\begin{equation}
r\frac{u_0'(r)}{u_0(r)}=\frac{r}{r-a}.
\end{equation}
This needs to be equal to the logarithmic derivative of Eq.~(\ref{eq:l=0_radial_solution}), $kr \cot(kr+\delta_0)$. Hence,
\begin{equation}
kr \cot(kr+\delta_0)=\frac{r}{r-a}.
\end{equation}
In the limit $k\to 0$, and also setting $r=0$, we have 
\begin{equation}\label{eq:scattering_length}
\lim_{k\to 0} k \cot\delta_0(k) = -\frac{1}{a},
\end{equation}
where $a$ is called scattering length~\cite{Sakurai2014}. Let us consider the implications of this result. In the previous section, we reduced the scattering problem to calculating the phase shifts $\delta_l(k)$. Here we are considering only low-energy phenomena, to which we argued that the $l=0$ component is the dominant one. However, Eq.~(\ref{eq:scattering_length}) reduces the problem even further: in the zero-energy limit, a single number, the scattering length, encodes all the information we need about scattering.

\begin{figure}[!htb] 
    \centering
    \begin{minipage}{0.5\textwidth}
        \centering
        \subfloat[An attractive potential that is not strong enough to produce a bound state. In this case, $a<0$ because we need to extrapolate the radial function to negative values to intercept the $r$-axis. \label{fig:fig:scattering_intercept_a}]{\includegraphics[width=0.8\textwidth]{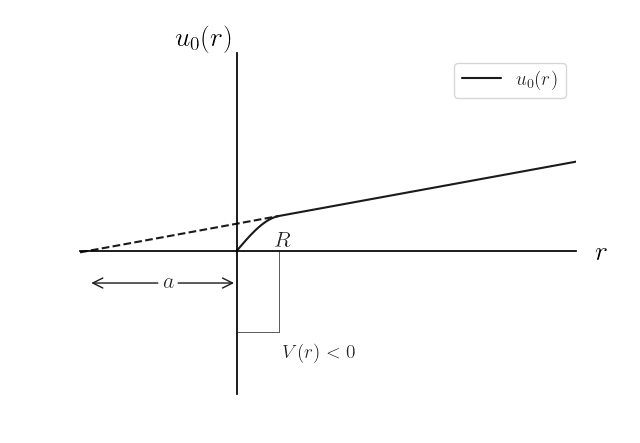}}
    \end{minipage}\hfill
    \begin{minipage}{0.5\textwidth}
        \centering
        \subfloat[A stronger attractive potential produces a bound state, and $a>0$. \label{fig:scattering_intercept_b}]{\includegraphics[width=0.8\textwidth]{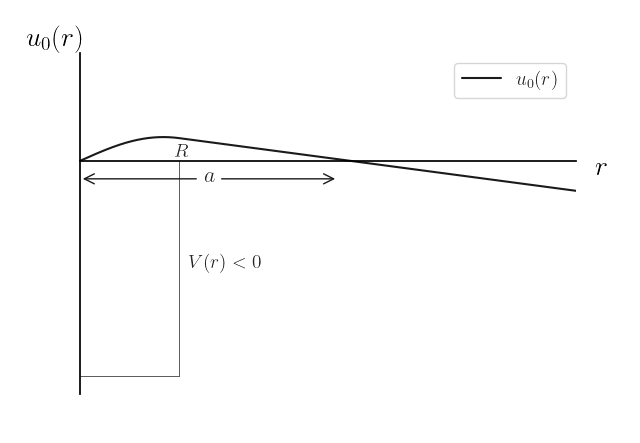}}
    \end{minipage}\hfill
    \begin{minipage}{0.5\textwidth}
        \centering
        \subfloat[For a repulsive potential, we always have $a>0$. \label{fig:scattering_intercept_c}]{\includegraphics[width=0.8\textwidth]{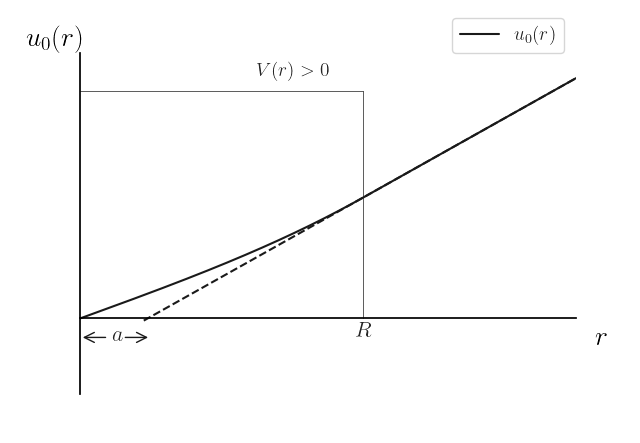}}
    \end{minipage}
        \caption{Geometrical interpretation of the scattering length $a$ based on three possibilities.}
    \label{fig:scattering_intercept}
\end{figure}

The scattering length has, as its name suggests, dimension of length. However, it may differ by orders of magnitude from the range $R$ of the potential, as we will see in several examples. A geometrical interpretation is possible if we choose $c=-1/a$ in Eq.~(\ref{eq:line}) (remember that the normalization always cancels out when we work with ratios such as $u_0'/u_0$),
\begin{equation}\label{eq:outside_zero_energy_solution}
    u_0(r>R) = 1- \frac{r}{a}.
\end{equation}
We see that the scattering length is simply the intercept of the outside wave function, or its extrapolation, as illustrated in Fig.~\ref{fig:scattering_intercept}. Finally, the scattering amplitude for $l=0$, Eq.~(\ref{eq:fl}), in the low-energy limit is 
\begin{equation}
f_0(k) = \lim_{k\to 0} \frac{1}{k\cot\delta_0-ik}  = -a.
\end{equation}

\subsection{The effective range}
\label{sec:effective_range}

Equation~(\ref{eq:scattering_length}) and the concept of scattering length are remarkable. However, the underlying $kR\ll 1$ assumption makes them good approximations to physical situations only when the energy or the range of the potential goes to zero. We might wonder if we can modify Eq.~(\ref{eq:scattering_length}) to consider a finite but small value of $kR$. In other words, we would like to express $k \cot\delta_0(k)$ as a series in powers of $k$. We already know that the first term is $-1/a$, so the task becomes computing the next.

We follow the procedure outlined in Ref.~\cite{Madsen2002}. First, let us choose a different normalization for $u_0(r)$ in Eq.~(\ref{eq:l=0_radial_solution}),
\begin{equation}
\label{eq:ur_norm}
u_0(r)=\cot\delta_0(k) \sin(kr)+\cos(kr),
\end{equation}
and the reason for this choice will be apparent shortly.
Let us take the $l=0$ radial equation, Eq.~(\ref{eq:radial_equation}), for two different wave functions $u_{k_1}(r)$ and $u_{k_2}(r)$, labeled by their wave vectors $k_1=\sqrt{2mE_1}/\hbar$ and $k_2=\sqrt{2mE_2}/\hbar$, 
\begin{eqnarray}
u_{k_1}''(r) - U(r)u_{k_1}(r)+k_1^2u_{k_1}(r) = 0,\nonumber\\
u_{k_2}''(r) - U(r)u_{k_2}(r) +k_2^2u_{k_2}(r) = 0.
\end{eqnarray}
Next, we multiply the first equation by $u_{k_2}$ and the second by $u_{k_1}$ and take their difference,
\begin{equation}\label{eq:uk1-uk2_difference}
u_{k_1}''(r)u_{k_2}(r) - u_{k_1}(r)u_{k_2}''(r) = (k_2^2-k_1^2)u_{k_1}(r)u_{k_2}(r).
\end{equation}
We may write the left-hand side as
\begin{eqnarray}
    u_{k_1}''(r)u_{k_2}(r) - u_{k_1}(r)u_{k_2}''(r) = \nonumber\\
    \frac{d}{dr}\left[ u'_{k_1}(r)u_{k_2}(r)-u'_{k_2}(r)u_{k_1}(r)\right].
\end{eqnarray}
Now we integrate Eq.~(\ref{eq:uk1-uk2_difference}) from $0$ to $R$,
\begin{eqnarray}
\label{eq:diff_k1_k2}
\left[u'_{k_2}(r)u_{k_1}(r) - u'_{k_1}(r)u_{k_2}(r)\right]_0^R = \nonumber\\
(k_2^2-k_1^2)\int_0^R dr\, u_{k_1}(r)u_{k_2}(r).
\end{eqnarray}
The integral converges since $A_0(r)=u_0(r)/r$ is finite at the origin ($u_0(0) = 0$ independently of the energy).

Next, we repeat the same procedure for the free-particle ($U=0$) radial equation with solutions denoted by $g_{k_1}(r)$ and $g_{k_2}(r)$. The result is the same as Eq.~(\ref{eq:diff_k1_k2}) if we replace $u$ by $g$. Finally, we take the difference between this result and Eq.~(\ref{eq:diff_k1_k2}),
\begin{flalign}
\label{eq:diff_g_u}
&\left[g'_{k_2}(r)g_{k_1}(r) - g'_{k_1}(r)g_{k_2}(r)\right]_0^R & \nonumber\\
&- \left[u'_{k_2}(r)u_{k_1}(r) - u'_{k_1}(r)u_{k_2}(r)\right]_0^R= & \nonumber \\ 
& (k_2^2-k_1^2)\int_0^R dr \,[g_{k_1}(r)g_{k_2}(r)-u_{k_1}(r)u_{k_2}(r)]. & 
\end{flalign}
The functions $g_{k_1}$ and $g_{k_2}$ are given by Eq.~(\ref{eq:ur_norm}), and so are $u_{k_1}$ and $u_{k_2}$ for $r\geqslant R$. Then, Eq.~(\ref{eq:diff_g_u}) becomes
\begin{flalign}
& k_2\cot\delta_0(k_2)-k_1\cot\delta_0(k_1) = & \nonumber\\
& (k_2^2-k_1^2)\int_0^R dr \,[g_{k_1}(r)g_{k_2}(r)-u_{k_1}(r)u_{k_2}(r)]. &
\end{flalign}
If we take the limit $k_1 \to 0$, we can write $k_1\cot\delta_0(k_1)$ in terms of the scattering length, Eq.~(\ref{eq:scattering_length}),
\begin{equation}
k\cot\delta_0(k) = -\frac{1}{a}+k^2\int_0^R dr \,[g_{0}(r)g_{k}(r)-u_{0}(r)u_{k}(r)],
\end{equation}
where we dropped the subscript in $k_2$ for simplicity. We define the quantity
\begin{eqnarray}
\rho(k)\equiv 2 \int_0^R dr \,[g_{0}(r)g_{k}(r)-u_{0}(r)u_{k}(r)].
\end{eqnarray}
Finally, we have an expression for $k\cot\delta_0(k)$ at low-energies,
\begin{equation}\label{eq:shape_independent_equation}
k \cot\delta_0(k) = -\frac{1}{a} + \frac{1}{2}r_0k^2 + \mathcal{O}(k^4),
\end{equation}
where the term $r_0$ is referred to as ``effective range'' and is defined as
\begin{equation}\label{eq:effective-range}
r_0 \equiv \lim_{k\to 0} \rho(k)=2 \int_0^R dr \, [g_0^2(r) - u_0^2(r)],
\end{equation}
where $g_0(r)$ is calculated by taking $k\to 0$ in Eq.~(\ref{eq:ur_norm}). The result is exactly Eq.~(\ref{eq:outside_zero_energy_solution}), justifying the normalization choice.

Comparing Eqs.~(\ref{eq:scattering_length}) and (\ref{eq:shape_independent_equation}), we see that we were able to include the next term in the expansion, which is proportional to $k^2$. Only even powers of $k$ appear on the right-hand side of Eq.~(\ref{eq:shape_independent_equation}) because of the symmetry with respect to changing the sign of $k$: the phase shift would also change sign and, since $\cot(-x)=-\cot(x)$, the left-hand side is unchanged. Hence, the right-hand side can only have even functions of $k$. To compute the effective range, we only need two zero-energy solutions of the radial equation: the free particle, $g_0$, and the solution with the potential $V(r)$, $u_0$.

Equation~(\ref{eq:shape_independent_equation}) asserts that the $s$-wave scattering phase shift does not depend on the shape of the potential $V(r)$ at low energies. Instead, two potentials with different forms will produce the same phase shift if their scattering lengths and effective ranges are the same. For this reason, Eq.~(\ref{eq:shape_independent_equation}) is commonly referred to as \textit{shape-independent approximation}. Nevertheless, it is important to note the error order of $k^4$. At higher energies, outside the scope of this work, higher order terms become relevant, and Eq.~(\ref{eq:shape_independent_equation}) may not be appropriated.

\subsection{Bound states}
\label{sec:bound_states}

When we defined the phase shifts in terms of the ratio of the coefficients appearing in front of the outgoing and income spherical waves, Eq.~(\ref{eq:scattering_ratio}), we also introduced the amplitude $S_l(k)$. The analytic properties of $S_l(k)$ contain information about bound states, as we will see in the following.

Let us rewrite Eq.~(\ref{eq:asymptotic_hankel_wavefunction}) as
\begin{eqnarray}
\psi(r,\theta) \xrightarrow{\text{large }r} \frac{1}{(2\pi)^{3/2}}\sum_{l=0}^\infty \frac{(2l+1)}{2ik} P_l(\cos\theta) \times \nonumber\\
\left[S_{l}(k) \frac{e^{ikr}}{r} - \frac{e^{-i(kr-l\pi)}}{r}\right].
\end{eqnarray}
For $l=0$ and large distances, the radial wave function is proportional to
\begin{equation}
\label{eq:ratio}
S_{0}(k)\frac{e^{ikr}}{r}-\frac{e^{-ikr}}{r}.
\end{equation}

For an arbitrary finite-ranged potential $V$, the radial solution at $r>R$ for a bound state ($E<0$) obeys
\begin{equation}
u''(r) = -\frac{2mE}{\hbar^2}u(r) = \kappa^2u(r),\quad \kappa\equiv \frac{\sqrt{-2mE}}{\hbar}.
\end{equation}
The solution can be written as
\begin{equation}
u(r>R) = Ae^{\kappa r}+Be^{-\kappa r},   
\end{equation}
where $A$ and $B$ are constants. However, $u(r\to\infty)$ must be finite, so we set $A=0$. We conclude that the radial function for a bound state at large distances is
\begin{equation}
\label{eq:bound_large_r}
A(r)=\frac{u(r)}{r} \propto \frac{e^{-\kappa r}}{r} \text{ (large $r$)}.
\end{equation}

\begin{figure}[!htb]
    \centering
    \includegraphics[width=\linewidth]{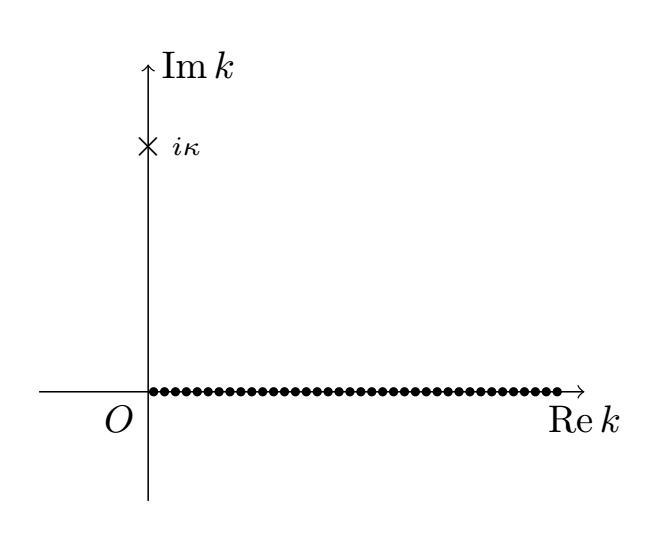}
    \caption{Complex $k$-plane. A bound state is represented by a pole in the $S$-matrix at $k=i\kappa$, while the scattering continuum corresponds to a real value of $k>0$.}
    \label{fig:s_matrix_pole}
\end{figure}

The existence of a bound state is related to the Schr\"odinger equation admitting a solution with a discrete $E<0$ energy. By looking at Eq.~(\ref{eq:ratio}), we may be tempted to substitute $k\to i\kappa$, with $k$ purely imaginary, to connect it to the bound state behavior of Eq.~(\ref{eq:bound_large_r}),
\begin{equation}
\frac{e^{ikr}}{r}=\frac{e^{i(i\kappa)r}}{r}=\frac{e^{-\kappa r}}{r}.
\end{equation}
However, the important difference is that the bound state is present even without the incident wave we have used for our scattering problem formulation. 
In Eq.~(\ref{eq:ratio}), $S_0(k)$ controls the ratio of the outgoing to the incoming wave. Hence, in the bound state case, we have only the outgoing component, meaning an infinite ratio: $S_0(k)\to\infty$. Since we are treating $S_0(k)$ as a function of a complex variable, this means that it has a pole at $k=i\kappa$. Hence, the conclusion is that a bound state appears as a pole in $S_0(k)$. It is convenient to represent the variable $k$ in the complex $k$-plane, i.e., where the abscissa is the real part of $k$, and the ordinate is the imaginary part of $k$. The pole at $k=i\kappa$ is a point on the $\operatorname{Im}(k)$ axis, while the scattering continuum is on the positive $\operatorname{Re}(k)$ axis, since $k>0$, as illustrated in Figure~\ref{fig:s_matrix_pole}.

Let us derive an expression for $S_0(k)$, so we can identify the $k=i\kappa$ pole.
Eqs.~(\ref{eq:fl}) and (\ref{eq:scattering_length}) yield
\begin{equation}
f_0(k) = \frac{1}{-1/a-ik}.
\end{equation}
The scattering amplitude is related to $S_0(k)$ through Eq.~(\ref{eq:partial_wave_amplitude-scattering_matrix_element}),
\begin{equation}
S_0(k)=1+2ikf_0(k)=\frac{-k-i/a}{k-i/a}.
\end{equation}
This expression has a pole at $k=i\kappa$ if we identify
\begin{equation}
\label{eq:kappa=1overa}
\kappa=\frac{1}{a}.
\end{equation}
Hence, in the zero-energy limit, the energy of a bound state and the scattering length are connected simply by
\begin{equation}
\label{eq:E_one_part}
E=-\frac{\hbar^2\kappa^2}{2m}=-\frac{\hbar^2}{2ma^2}.
\end{equation}
A single parameter originated from the potential determines the bound-state energy.

\subsection{Two-body scattering}\label{sec:two-body-scattering}

So far, we considered only the problem of a single particle being scattered by a finite-ranged potential $V(r)$ located at $r=0$. With a few modifications, we can use the results we obtained to describe two particles interacting through a pairwise potential which depends only on their spatial separation $r$.

Besides being separable in radial and angular coordinates, the Hamiltonian of a two-body system with a spherically symmetric potential is also separable in the center of mass (CM) and relative coordinates. The two-body Hamiltonian is given by
\begin{equation}
H = -\frac{\hbar^2}{2m_1} \nabla_{\mb{r}_1}^2 -\frac{\hbar^2}{2m_2}\nabla_{\mb{r}_2}^2  + V(\mb{r}_1-\mb{r}_2),  
\end{equation}
where the indices $1$ and $2$ refer to different particles.
We define the coordinates
\begin{eqnarray}
\mb{R}&=&\frac{m_1\mb{r}_1+m_2\mb{r}_2}{M},\nonumber\\
\mb{r}&=&\mb{r}_1-\mb{r}_2,
\end{eqnarray}
where $M=m_1+m_2$.
In this coordinate system, and using the fact that the potential depends only on $r = |\mb{r}|$,
\begin{eqnarray}
H &=& H_{\text{CM}} + H_r,\nonumber\\
H_{\rm CM} &=& -\frac{\hbar^2}{2M}\nabla_{\mb{R}}^2,\nonumber\\
H_r &=& - \frac{\hbar^2}{2m_r}\nabla_{\mb{r}}^2 + V(r),
\end{eqnarray}
where $m_r = m_1m_2/(m_1+m_2)$ is the reduced mass.
The CM motion satisfies the free-particle equation, so its behavior is trivial and only adds a constant to the total energy. However, the relative movement is non-trivial and contains all the physics due to the scattering of the two particles. Notice that the relative motion Hamiltonian is exactly the one we used to discuss the one particle problem if we substitute $m\to m_r$. Thus we can apply our previous results to two-body scattering.

\subsection{Some applications}

Before presenting the numerical procedure to calculate the scattering length and effective range of two-body potentials, it is instructive to show some examples of analytical calculations.

\subsubsection{Spherically symmetric finite well}\label{sec:finite_well}

For example, let us consider a two-body system with an interaction described by an attractive spherically symmetric finite well. It has a constant value inside the range $R$ and zero outside. We can write $V(r)$ as 
\begin{equation}\label{eq:square_well_potential}
V(r) = 
\begin{dcases}
-v_0\,\frac{\hbar^2}{m_rR^2}, \quad &\text{for } r<R,\\
0, \quad & \text{for } r>R,
\end{dcases}
\end{equation}
where $v_0>0$ is a dimensionless parameter related to the depth of the well. We generally want to tune the parameters $v_0$ and $R$ to achieve the physical conditions we are interested in studying. For a relatively shallow or short-ranged potential, we may only observe continuum scattering states ($E>0$), but increasing its depth or range may make it strong enough to produce a bound state ($E<0$).

Let us start with the $E>0$ case. For simplicity, let us consider the equal mass case, $m_1=m_2=m$. The $s$-wave ($l=0$) equation we need to solve is
\begin{equation}\label{eq:l=0_schrodinger}
\left(\frac{d^2}{dr^2} - \frac{2m_r}{\hbar^2} V(r) + \frac{2m_r}{\hbar^2}E \right)u(r) = 0 ,
\end{equation}
where $u(r) = rA(r)$. Writing the explicit forms inside and outside the range of the potential yields
\begin{align}
    &u''(r) + \left(k_0^2 +k^2\right)u(r) = 0 &\text{for }r<R, \nonumber \\
    &u''(r) +k^2u(r)  = 0 &\text{for }r>R,
\end{align}
where $k^2 \equiv 2m_rE/\hbar^2$ and $k_0^2\equiv 2v_0/R^2$. In the region $r<R$, the solution may be written as
\begin{equation}
u(r) = A\sin\left(\sqrt{k^2+k_0^2}\ r\right) + B\cos\left(\sqrt{k^2+k_0^2}\ r\right).
\end{equation}
Since $u(0)=0$, because the radial solution $A(r)=u(r)/r$ is regular at the origin, we set $B=0$. In the region $r>R$, $u(r)$ is given by Eq.~(\ref{eq:ur_norm}). Hence, the solution is of the form
\begin{equation}\label{eq:well_scat_solutions}
u(r) = 
\begin{cases}
A\sin\left(\sqrt{k^2+k_0^2}\ r\right) &\text{for } r<R,\\
\cot\delta_0(k) \sin(kr)+\cos(kr) &\text{for } r>R,
\end{cases}
\end{equation}
where $\delta_0(k)$ is the phase shift. Now we equate the logarithmic derivative at $r=R^{-}$ and $r=R^{+}$,
\begin{equation}
    \left[r\frac{u'(r)}{u(r)}\right]_{r=R^{-}} = \left[r\frac{u'(r)}{u(r)}\right]_{r=R^{+}},
\end{equation}
which yields,
\begin{eqnarray}\label{eq:well_log_derivative}
\frac{\sqrt{k^2+k_0^2}\cos\left(\sqrt{k^2+k_0^2}\ R\right)}{\sin\left(\sqrt{k^2+k_0^2}\ R\right)} = \nonumber\\
\frac{k\cot\delta_0(k)\cos(kR)-k\sin(kR)}{\cot\delta_0(k) \sin(kR)+\cos(kR)}.
\end{eqnarray}
Without any approximation, we may solve for the phase shift $\delta_0(k)$,
\begin{equation}\label{eq:well_phase-shift}
\delta_0(k) = -kR + \arctan\left[\frac{k\tan\left(\sqrt{k^2+k_0^2}\,\,R\right)}{\sqrt{k^2+k_0^2}}\right].
\end{equation}

To calculate the scattering length $a$, we need to take the $k\to 0$ limit as prescribed in Eq.~(\ref{eq:scattering_length}). One straightforward way to do this is to rearrange Eq.~(\ref{eq:well_log_derivative}) so that we have factors of $k\cot\delta_0(k)$. It is important to keep track of the orders employed in the approximation: Eq.~(\ref{eq:scattering_length}) ignores $\mathcal{O}(k^2)$ terms. Hence, to be consistent, we need to use the following approximations,
\begin{eqnarray}
\cos(kR) &=& 1 + \mathcal{O}(k^2),\nonumber\\
\sin(kR) &=& kR + \mathcal{O}(k^3).
\end{eqnarray}
Performing these approximations, Eq.~(\ref{eq:well_log_derivative}) becomes
\begin{equation}
\sqrt{k_0^2}\cot\left(\sqrt{k_0^2}\ R\right) = \frac{-1/a}{-R/a+1}.
\end{equation}
Finally, solving for $a$,
\begin{equation}\label{eq:well-scattering_length}
a =R-\frac{\tan\left(\sqrt{k_0^2} R\right)}{\sqrt{k_0^2}}=R\left(1 - \frac{\tan\left(\sqrt{2 v_0}\right)}{\sqrt{2 v_0}}\right),
\end{equation}
where we used $k_0^2=2v_0/R^2$ in the last step. Equation~(\ref{eq:well-scattering_length}) makes it apparent that the scattering length depends only on the parameters of the potential, its depth $v_0$ and range $R$. Further inspection of the equation leads us to conclude that the scattering length diverges for specific values of $v_0$. As we will see in the following, these are related to the appearance of bound states.

We could repeat the above procedure if we want to consider the $E<0$ cases. However, suppose we recall the discussion of bound states in Sec.~\ref{sec:bound_states}. In that case, we conclude that we only need to replace $k=i\kappa$ in the solution inside the range of the potential and that the solution outside is proportional to $e^{-\kappa r}$. Thus, the equivalent of Eq.~(\ref{eq:well_scat_solutions}) for $E<0$ is
\begin{equation}
u(r) = 
\begin{cases}
A'\sin\left(\sqrt{k_0^2-\kappa^2}\ r\right) &\text{for } r<R,\\
B'e^{-\kappa r} &\text{for } r>R,
\end{cases}
\end{equation}
where $A'$ and $B'$ are constants. They can be determined by matching $u(r)$ at $r=R$ and normalizing the radial solution. Next, we match the logarithmic derivatives at $r = R$,
\begin{equation}
\frac{\sqrt{k_0^2-\kappa^2}\cos\left(\sqrt{k_0^2-\kappa^2}\ R\right)}{\sin\left(\sqrt{k_0^2-\kappa^2}\ R\right)} = \frac{-\kappa e^{-\kappa R}}{e^{-\kappa R}}.
\end{equation}
After some manipulations,
\begin{equation}\label{eq:transcendetal}
\tan\left(\sqrt{k_0^2-\kappa^2}\ R\right) +\frac{\sqrt{k_0^2-\kappa^2}}{\kappa} = 0.
\end{equation}
This is a transcendental equation that shows where the bound-state energies are located. It cannot be solved analytically, although obtaining numerical solutions is straightforward. Nevertheless, the term $\sqrt{k_0^2-\kappa^2}/\kappa$ is always positive. The only way for Eq.~(\ref{eq:transcendetal}) to be valid is if $\tan(\sqrt{k_0^2-\kappa^2}\ R)$ is negative. That is to say,
\begin{equation}
\label{eq:ineq_well}
\frac{\pi}{2} + n\pi < \sqrt{k_0^2-\kappa^2}\ R <\pi + n\pi,
\end{equation}
where $n$ is an integer. Since $\sqrt{k_0^2-\kappa^2}\ R$ is always positive, $n=0,1,...$. The first bound state is $n=0$. Thus,
\begin{equation}
\label{eq:ineq_neq0_well}
\frac{\pi}{2R} < \sqrt{k_0^2-\kappa^2} <\frac{\pi}{R}.
\end{equation}
Since $k_0>\sqrt{k_0^2-\kappa^2}$ then, substituting $k_0=\sqrt{2v_0}/R$ and using Eq.~(\ref{eq:ineq_neq0_well}), we have
\begin{equation}
\label{eq:bound_state_threshold}
v_0 > \frac{\pi^2}{8}.
\end{equation}
This result shows that there are no bound states if $v_0$ is not above a certain threshold value.

Now we have everything we need to connect the appearance of bound states and the scattering length. Equation~(\ref{eq:well-scattering_length}) expresses the scattering length $a$ in terms of $v_0$. We know that $\tan x$ diverges for $x=\pi/2$, which implies that $a$ diverges for $v_0=\pi^2/8$. This value is exactly the threshold we found for the appearance of a bound state, Eq.~(\ref{eq:bound_state_threshold}). The conclusion is that the scattering length diverges when a bound state appears. However, it depends on which side we approach since
\begin{eqnarray}
\lim_{x\to \frac{\pi}{2}^-} \tan x &=& +\infty, \nonumber\\
\lim_{x\to \frac{\pi}{2}^+} \tan x &=& -\infty.
\end{eqnarray}

\begin{figure}[!htb]
\centering
\includegraphics[width=\linewidth]{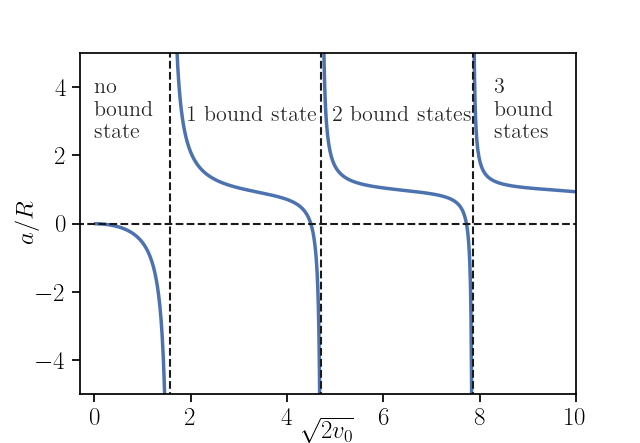}
\caption{Behavior of the scattering length $a$ as function of $\sqrt{2v_0}$ for the spherical well, Eq.~(\ref{eq:well-scattering_length}). For $\sqrt{2v_0}= \pi/2 + n\pi$ ($n =0,1,2,...$), $a$ diverges. This is related to the potential admitting an additional bound state at those points.}
\label{fig:well-scattering-length}
\end{figure}

Let us consider the situation where we fixed the range $R$ of the potential and start with a very shallow well. Then $a<0$, and no bound states are supported. If we increase the depth of the well, $|a|$ also increases to the point we are at the threshold for a bound state to appear. At this point, $|a|\to\infty$ and a bound state with $E=0$ appears. If we continue to increase $v_0$, then $a$ changes sign, and it will decrease until the formation of a second bound state. This situation is illustrated in Fig.~\ref{fig:well-scattering-length}. We centered our discussion around the appearance of the first bound state, but Eq.~(\ref{eq:ineq_well}) implies that this behavior will happen whenever the system is at the threshold of forming an additional bound state. Figure~\ref{fig:well-scattering-length} shows this since the points where $|a|\to \infty$ are exactly where $\sqrt{2v_0} = \pi/2 + n\pi$.

The advantage of the low-energy scattering formulation is that we may study the behavior of the bound states of the potentials without explicitly solving Schr\"odinger's equation. With the benefit of hindsight, Eq.~(\ref{eq:E_one_part}) had already told us that $|a|\to\infty$ corresponds to the formation of a new bound state since that, at the threshold, the state must have $E=0$.

\begin{figure}[!htb]
\centering
\includegraphics[width=\linewidth]{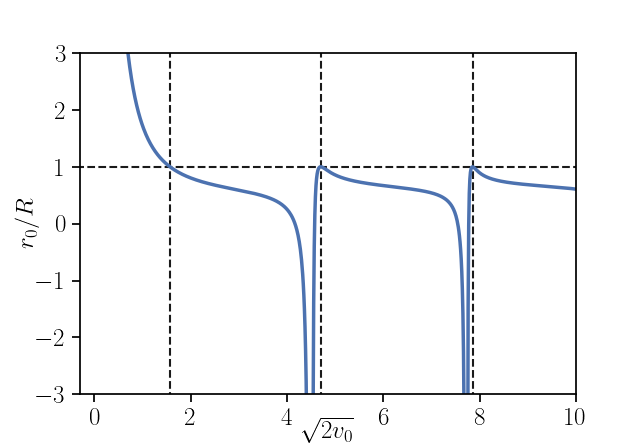}
\caption{Behavior of the effective range $r_0$ as function of $\sqrt{2v_0}$ for the spherical well, Eq.~(\ref{eq:well-effective-range}). The vertical dashed lines denote where the scattering length diverges, $|a| \to \infty$, corresponding to an effective range equal to the range of the potential, $r_0/R=1$.}
\label{fig:well-effective-range}
\end{figure}

\renewcommand{\arraystretch}{1.15}
\begin{table*}[!htb]
    \centering
    \caption{Summary of the low-energy properties of two physical systems: the $^4$He dimer and the deuteron. The values of the scattering length $a$, effective range $r_0$, and bound-state energy $E$ are taken from the indicated references. The zero-range and finite-range approximations, $E_{zr}$ and $E_{fr}$, were calculated using Eqs.~(\ref{eq:Ezr}) and (\ref{eq:Efr}).}
    \begin{tabular}{ccccccc}
    \hline\hline
    System   & $a$ & $r_0$ & $E$ & Ref. & $E_{zr}$ & $E_{fr}$\\
    \hline
    $^4$He dimer & 90.4 \AA & 8.0 \AA & $-$1.62 mK & \cite{Cencek2012} & $-$1.48 mK & $-$1.63 mK\\
    Deuteron & 5.4112 fm & 1.7436 fm & $-$2.224 MeV & \cite{Hackenburg2006} & $-$1.416 MeV & $-$2.223 MeV  \\
    \hline\hline
    \end{tabular}
    \label{tab:zr_fr}
\end{table*}

Finally, to complete our analysis of the spherical well, we must compute the effective range as defined in Eq.~(\ref{eq:effective-range}). Notice that the normalization of the solution outside the range of the potential in Eq.~(\ref{eq:well_scat_solutions}) agrees with the one we used while deriving the effective range expansion. To determine the constant $A$ we impose the continuity of $u(r)$ at $r=R$,
\begin{equation}
A = \frac{\cot\delta_0(k) \sin(kR)+\cos(kR)}{\sin(\sqrt{k^2+k_0^2}R)}.
\end{equation}
The normalized solution is written as 
\begin{flalign}\label{eq:well_norm_scat_solutions}
&u(r) =&\nonumber\\
&
\begin{cases}
    \frac{\cot\delta_0(k)\sin(kR)+\cos(kR)}{\sin(\sqrt{k^2+k_0^2}R)}
\sin\left(\sqrt{k^2+k_0^2}\ r\right) &\text{for } r<R,\nonumber \\
     \cot\delta_0(k)\sin(kr)+\cos(kr) &\text{for } r>R.
\end{cases}&\\
\end{flalign}
The effective range, Eq.~(\ref{eq:effective-range}), is defined in the $k\to 0$ limit of $u(r)$,
\begin{equation}\label{eq:well_k_lim_scat_solutions}
u(r) = 
\begin{cases}
    \frac{\left(1-R/a\right)}{\sin(k_0 R)}
\sin(k_0 r) &\text{for } r<R,\\
     1-r/a  &\text{for } r>R.
\end{cases}
\end{equation}
Then, the effective range is given by the integral
\begin{equation}
r_0 = 2\int_0^R dr\,\left[\left(1-\frac{r}{a}\right)^2-\left(1-\frac{R}{a}\right)^2\frac{\sin^2(k_0 r)}{\sin^2(k_0 R)}\right].
\end{equation}
After computing the integral, we use Eq.~(\ref{eq:well-scattering_length}) to replace $a$ in favor of $R$ and $k_0$. The result is
\begin{eqnarray}\label{eq:well-effective-range}
r_0 = R\left(1-\frac{1}{3} \left(\frac{k_0 R}{\tan(k_0 R)-k_0 R}\right)^2 +\right.\nonumber\\
\left.\frac{1}{k_0 R\tan(k_0 R) - (k_0 R)^2} \right),
\end{eqnarray}
which shows that $r_0$ is also dependent on parameters of the potential only. We illustrate the behavior of $r_0$ in Fig.~\ref{fig:well-effective-range}. Another characteristic is that the range of the potential and the effective range are equal ($r_0=R$) only when the scattering length diverges. This stresses that even for the case of a spherical well when there is no doubt that the range of the potential is $R$, a rigorous analysis should be carried out in terms of the effective range if we want to capture the low-energy properties of the system.

\subsubsection{Zero-range and finite-range approximations}
\label{sec:approximations}

Equation~(\ref{eq:kappa=1overa}) has an immediate application when dealing with two-body bound states in the zero-energy limit. The energy can be readily calculated as follows:
\begin{equation}
\label{eq:Ezr}
E_{zr}=-\frac{\hbar^2\kappa^2}{2 m_r}=-\frac{\hbar^2}{2 m_r a^2},
\end{equation}
where we chose the subscript to denote the zero range, i.e., in deriving this result, $r_0$ was taken to be zero. If we want to modify the equation above to include finite-range corrections, we have to combine these results with Eq.~(\ref{eq:shape_independent_equation}) and make the substitution $k\to i\kappa$, which yields~\cite{Jamieson1995,Janzen1995}
\begin{equation}
\frac{1}{a}=\kappa-\frac{1}{2}r_0\kappa^2.
\end{equation}
We can solve this quadratic equation for $\kappa$, choose the appropriate root, and use it to compute the bound state energy,
\begin{equation}
\label{eq:Efr}
E_{fr}=-\frac{\hbar^2\kappa^2}{2 m_r}=-\frac{\hbar^2}{2m_r r_0^2}\left(1-\sqrt{1-\frac{2r_0}{a}}\right)^2,
\end{equation}
where the subscript $fr$ stands for finite-range.

Equations~(\ref{eq:Ezr}) and (\ref{eq:Efr}) seem very powerful in the sense that we only need one or two zero-energy parameters to compute the energy of a bound state, besides the reduced mass. However, the question of whether this model produces sensible results in realistic settings remains. At this point, applying the zero-range and finite-range results to physical systems is illustrative. To clarify, we do not compute $a$ and $r_0$ for the examples below; we just take values from the literature. Table~\ref{tab:zr_fr} contains a summary of the parameters we employed and the results obtained.

We present two examples; the first is the bound state of two $^4$He atoms, called helium dimer. There are many realistic pairwise potentials for $^4$He. Let us consider the one of Ref.~\cite{Cencek2012}. It contains adiabatic, relativistic, and quantum electrodynamics interactions. Explaining the details of the potential is beyond the scope of this work, but we only need three values from this reference to illustrate how useful the effective-range approximation is. The authors reported that their potential produces a dimer energy of $E_d=-1.62$ mK, a scattering length of $a=90.4$ \AA, and an effective range of $r_0=8.0$ \AA.

Substituting $a$, and the mass of an $^4$He atom, in Eq.~(\ref{eq:Ezr}) yields $E_{zr}\approx -1.48$ mK, such that $E_{zr}/E_d\approx 0.92$. Let us pause to appreciate this result. We estimated the energy of a bound-state using only one parameter calculated from a zero-energy theory, and it differs only 8\% from the full solution of the Schr\"odinger equation. If we include the finite-range correction, Eq.~(\ref{eq:Efr}), then the result is $E_{fr}=-1.63$ mK, which is correct within 1\%. Both the zero- and finite-range results successfully describe the physical system because $k R \sim 0.1$ in this case.

The second example we chose is from nuclear physics. The deuteron (the nucleus of deuterium) is the only bound state of two nucleons, and it is formed by a proton and a neutron. The details of nuclear interaction are much more complicated than what we present here. For example, the deuteron has a quadrupole moment due to a $l=2$ partial-wave component, which goes beyond $s$-wave ($l=0$) scattering. However, this component is small, and we will give an $s$-wave treatment to the deuteron.

Reference~\cite{Hackenburg2006} reports the values $a=5.4112$ fm and $r_0=1.7436$ fm for the proton-neutron scattering length and effective range, respectively. Unlike the dimer case, where the particles are identical, the proton and neutron masses are different, although this difference is small. The reduced mass is then $m_r=m_p m_n/(m_p+m_n)$. We employed the values of the proton and neutron masses from the 2018 CODATA recommended values~\cite{CODATA}.

The experimental energy of the deuteron is $E_{d}=-2.224$ MeV. Substituting the appropriate values into Eq.~(\ref{eq:Ezr}) yields $E_{zr}=-1.416$ MeV, meaning that the zero-energy theory only accounts for $\approx$ 64\% of the deuteron energy. However, if we include the finite-range correction, Eq.~(\ref{eq:Efr}), then the energy is $E_{fr}=-2.223$ MeV, practically coinciding with the experimental value. Unlike the helium dimer, the range of the potential needed to be taken into account to reproduce the physical system because $kR\sim 0.4$ for the deuteron is larger than in the $^4$He case.

We should emphasize that the scales are very different in both examples. The $^4$He dimer, in the realm of atomic physics, is in a spatial scale of \AA ($10^{-10}$ m), and the energy is of the order of $10^{-7}$ eV. For the deuteron, the lengths are in the femtometer ($10^{-15}$ m) scale, while the energy is of a few MeV ($10^6$ eV). This exemplifies how universal are these low-energy scattering results.

\section{Numerical procedure}

\label{sec:numerical-procedure}

Analytical expressions for low-energy scattering parameters are only available for a few potentials. Even in those cases, the calculations may be cumbersome, as we saw in Sec.~\ref{sec:finite_well} for one of the simplest potentials, the spherical well. In general, we need to calculate $a$ and $r_0$ numerically.

\subsection{Numerical solutions of Schr\"odinger's equation}
\label{sec:numerical_sch}

\subsubsection{Second-order central difference}
\label{sec:central_diff}

We need two main ingredients to compute the low-energy scattering parameters, the radial wave function inside and outside the range of the potential. The latter has an analytical expression, but the former can only be computed numerically for most potentials. The radial equation for $u(r)=r A(r)$ resembles the one-dimensional Schr\"odinger's equation, so we can employ any of the various numerical methods that are available in the literature~\cite{Giordano2005}.

A common strategy in computational physics is to consider the function $u(r)$ in a discrete set of points $r_i=i\Delta r$, where $i$ is an integer ranging from $i=0,..., N$ and $\Delta r$ is a small quantity. Then, our goal becomes finding the value of $u(r_i)$ for every $i$. First, writing the expressions for the first and second numerical derivatives in this scenario is useful. For that, let us consider two Taylor expansions of $u(r)$ around the points $r\pm\Delta r$, 
\begin{flalign}
& u(r+\Delta r)=u(r)+(\Delta r)u'(r)+\frac{(\Delta r)^2}{2}u''(r) &\nonumber\\
& +\frac{(\Delta r)^3}{6}u'''(r)+\cdots, &\nonumber\\
& u(r-\Delta r)=u(r)-(\Delta r)u'(r)+\frac{(\Delta r)^2}{2}u''(r) &\nonumber\\
& -\frac{(\Delta r)^3}{6}u'''(r)+\cdots &
\end{flalign}
The difference of the two Taylor expansions yields an expression for the first derivative, while their sum results in the second derivative,
\begin{eqnarray}
\label{eq:first_der}
\frac{d u}{dr}\bigg|_{r=r_i} &\approx& \frac{u_{i+1}-u_{i-1}}{2\Delta r},\\
\label{eq:second_der}
\frac{d^2 u}{dr^2}\bigg|_{r=r_i} &\approx& \frac{u_{i+1}-2u_i+u_{i-1}}{(\Delta r)^2},
\end{eqnarray}
where we introduced a compact notation $u(r_i)\equiv u_i$ and, since $\Delta r$ is small, we neglected terms of $\mathcal{O}[(\Delta r)^3]$. These expressions tell us that if we want to calculate the numerical derivative at some point $r_i$, we need to know the value of the function at the neighboring sites; for the second derivative, we also need the value of the function at the site $r_i$. Since the expression for the first derivative is symmetric with respect to $i$, it is called the central difference. The same is true for Eq.~(\ref{eq:second_der}), called the second-order central difference. There are other numerical approximations to numerical differentiation, each with its advantages and drawbacks \cite{Press1996}.

Substituting Eq.~(\ref{eq:second_der}) into the $s$-wave zero-energy radial equation, we have
\begin{equation}
\label{eq:d2u_sym}
u_{i+1}=2 u_i - u_{i-1}+\frac{2m_r(\Delta r)^2}{\hbar^2}V(r_i) u_i.
\end{equation}
This equation tells us that if we know the value of the radial solution for two consecutive points, at $r_{i-1}$ and $r_i$, we can calculate the value for the next point $u_{i+1}$.

We know that $u(0)=0$; hence we need to know $u(\Delta r)$ to begin the calculation. For attractive potentials, $u(\Delta r)$ has a non-zero value. If we want to find a solution without worrying about the normalization, we can set $u(\Delta r)=1$. In other words, Schr\"odinger's equation is linear: if some $u(r)$ is a solution, then $C u(r)$ is also a solution, with $C$ constant.

Now we have everything to write a program that solves $u(r)$ inside the range $R$ of the potential $V(r)$. For reasons that will be apparent when we compute the scattering length, it is helpful to find the solution up to $R+\Delta r$. The inputs are typically the number of points $N$ (or the discretization $\Delta r$), and the parameters of the potential (in the case of the spherical well, Eq.~\ref{eq:square_well_potential}, $v_0$ and $R$). Once we know the number of points $N$, we may initialize an array of the desired dimension to hold the values of $u_i$. We also need to write a function that returns the value of $V(r_i)$ given a position $r_i$. Once these functionalities are implemented, the following algorithm will provide the values of $u_i$ for the whole interval:
\begin{enumerate}
    \item Set $u_0=0$, $u_1=1$, and $i=1$.
    \item Use Eq.(\ref{eq:d2u_sym}) to compute $u_{i+1}$.
    \item If $r_i\geqslant R+\Delta r$, stop. Else, increment $i$ by one.
    \item Go to step 2.
\end{enumerate}

\subsubsection{Numerov's method}

Equation~(\ref{eq:second_der}) is one possible discretization for a numerical second derivative. There are other alternatives if we want to improve the precision of our algorithm. For example, Numerov's method is a numerical technique capable of solving differential equations of second order when the first-order term is not present~\cite{Caruso2014}. It applies to equations of the form
\begin{equation}
\label{eq:numerov_yx}
\frac{d^2 y}{dx^2} = - \xi(x) y(x) + s(x).
\end{equation}
The method provides a relation between $y_i\equiv y(x_i)$ at three equally spaced consecutive points ($x_{i-1}$, $x_i$, and $x_{i+1}$),
\begin{flalign}
\label{eq:numerov_sol}
& y_{i+1}= \frac{1}{\left(1 + \frac{(\Delta x)^2}{12} \xi_{i+1}\right)}\Biggl\{ &\nonumber\\
& 2 y_i \left(1 - \frac{5 (\Delta x)^2}{12} \xi_i\right) - y_{i-1} \left(1 + \frac{(\Delta x)^2}{12} \xi_{i-1}\right) &\nonumber\\
&  + \frac{(\Delta x)^2}{12} (s_{i+1} + 10 s_i + s_{i-1}) \Biggr\} + \mathcal{O}[(\Delta x)^6],&
\end{flalign}
where $\Delta x$ is the spacing between the points, $\xi_i\equiv \xi(x_i)$, and $s_i\equiv s(x_i)$.

The appeal of this method to our case is immediate. The $s$-wave zero-energy radial equation is of the form of Eq.~(\ref{eq:numerov_yx}) with $y\to u$, $x\to r$, $s=0$, and
\begin{equation}
\xi(r)=-\frac{2m_r}{\hbar^2}V(r).
\end{equation}
The algorithm presented in Sec.~\ref{sec:central_diff} is mostly unchanged if we use Numerov's method instead of the second-order central difference. The key change is that substituting Eq.~(\ref{eq:d2u_sym}) by (\ref{eq:numerov_sol}) increases the precision without significant technical complications.

\subsubsection{Dimensionless quantities}

Schr\"odinger's equation contains relatively small quantities. Planck's reduced constant is $\hbar \sim 10^{-34}$ J s (or $\sim 10^{-15}$ eV s), while the typical masses, length, and energy scales are also small. This does not affect analytical calculations since the numerical values are commonly substituted at the end of the procedure. However, computers deal with real numbers differently. In computing, floating-point representation is an approximation that allows real numbers to be stored using an integer with a fixed precision, called the significand, scaled by an integer exponent of a fixed base. The number of bits dedicated to the significand and exponent determines the accuracy and range of the floating-point numbers that can be represented. However, due to the intrinsic limitations of representing real numbers in this format, rounding errors and other inaccuracies can occur, affecting the accuracy of computations involving floating-point numbers~\cite{Press1996}.

For these reasons, we would like to work within our program with quantities of the order of one (or close to it) so that we do not have to deal with very large or very small numbers. At the end of the calculation, once the solution is found, we may substitute the physical constants to compute the values for a particular system of interest. The procedure we describe here is equivalent to what is commonly called ``set $\hbar=m_r=1$'', but we provide a step-by-step derivation so that it is more clear how to recover the units at the end of the calculation.

Let us choose a length scale $\ell$. The convenient value of $\ell$ depends on the system under study; for atomic physics, it may be 1 \AA; for nuclear physics, we may use 1 fm or any other length scale that makes sense for a particular problem. In this section, we denote dimensionless quantities with bars over them. Then, the scaled distance
\begin{equation}
\label{eq:ell}
\bar{r}=\frac{r}{\ell}
\end{equation}
is dimensionless. The radial function $u(r)$ has units of [length]$^{-1/2}$ (a straightforward way to see this is to write the integral corresponding to its normalization). Hence the transformation to make it dimensionless is
\begin{equation}
\label{eq:dimensionless_u}
\bar{u}(\bar{r})=\frac{u(r)}{\ell^{-1/2}}.
\end{equation}
Equation~(\ref{eq:ell}) implies that $d^2/dr^2=(1/\ell^2)d^2/d\bar{r}^2$, so that the $s$-wave zero-energy radial equation becomes
\begin{equation}
-\frac{\hbar^2}{2m_r\ell^2}\frac{d^2\bar{u}}{d\bar{r}^2}+V(\bar{r})\bar{u} = 0.
\end{equation}
The quantity
\begin{equation}
\label{eq:energy_scale}
\epsilon=\frac{\hbar^2}{m_r\ell^2}
\end{equation}
has units of energy, so we may make the potential dimensionless by considering $\bar{V}=V/\epsilon$. Then, the equation we want to solve becomes
\begin{equation}
\label{eq:dimensionless}
-\frac{1}{2}\frac{d^2\bar{u}}{d\bar{r}^2}+\bar{V}(\bar{r})\bar{u} = 0.
\end{equation}
The net result is the same as ``set $\hbar=m_r=1$'', but now it is clear what to do. We implement and solve Eq.~(\ref{eq:dimensionless}) in our program. After we obtained the results in this dimensionless fashion, we use Eqs.~(\ref{eq:ell}), (\ref{eq:dimensionless_u}), and (\ref{eq:energy_scale}) to recover the solutions with the desired units.

\subsection{Scattering length and effective range}

After following the numerical procedure outlined in Sec.~\ref{sec:numerical_sch}, we should have the solution to the  $s$-wave zero-energy radial equation for the desired potential. That is, we calculated $u_i$ in the interval $r=0$ to $R+\Delta r$ at points $r_i$ equally spaced by $\Delta r$.

Equation~(\ref{eq:outside_zero_energy_solution}) is the low-energy ($k\to 0$) behavior of the radial solution outside the range of the potential, which we denote by $g_0(r)$ to avoid confusion. Then, we have the analytical expression,
\begin{equation}
\label{eq:gp_g}
\frac{g_0'(r)}{g_0(r)}=\frac{1}{r-a} \qquad \text{for } r>R.
\end{equation}
On the other hand, we can compute the same quantity at $r=R$ using the numerical solution we obtained. The numerical derivative $u'_{\rm num}$ can be calculated using Eq.~(\ref{eq:first_der}),
\begin{equation}
u'_{\rm num}(R)=\left.\frac{du(r)}{dr}\right|_{r=R} = \frac{u(R+\Delta r)-u(R-\Delta r)}{2\Delta r}.
\end{equation}
Now it becomes apparent why we included the point $R+\Delta r$ in the interval where we find $u(r)$ numerically. Finally, $u'_{\rm num}(R)/u_{\rm num}(R)$ must match Eq.~(\ref{eq:gp_g}) at $r=R$ (their logarithmic derivatives must be equal at this point), which gives us the expression
\begin{equation}
\label{eq:numerical_a}
a=R-\frac{2\Delta r\ u(R)}{u(R+\Delta r)-u(R-\Delta r)}.
\end{equation}
This expression relates the scattering length and our numerical solution.

Equation~(\ref{eq:numerical_a}) depends on the ratio of radial solutions, so we ignored its normalization. In other words, if we multiply all $u_i$ by a constant $C$, Eq.~(\ref{eq:numerical_a}) is unaltered. However, as we saw in Sec.~\ref{sec:effective_range}, the expression we derived for the effective range assumes a particular normalization choice. The constant $C$ is obtained by matching the value of the numerically-obtained radial function and the analytical solution at $r=R$, that is $C u(R)=g(R)$,
\begin{equation}
C=\frac{g(R)}{u(R)}=\frac{\left(1-R/a\right)}{u(R)}.
\end{equation}
So, in the following, we assume that all the $u_i$ have been multiplied by this constant $C$.

Finally, what is left is to compute the integral that defines the effective range, Eq.~(\ref{eq:effective-range}), numerically. Although the $V=0$ radial solution, denoted by $g(r)$, has an analytical expression, it is useful to discretize it at the same points where we determined $u_i$ numerically, $g_i \equiv g(r_i)$. Then, the task becomes computing an integral of the form
\begin{equation}
I=\int_{x_1}^{x_N} f(x) dx
\end{equation}
when the function $f(x)$ is known only at a discrete set of equally spaced points, $f(x_i)\equiv f_i$, where $i=1,2,3,...,N$.
This is a well-known problem in numerical calculus, and many methods accomplish the task~\cite{Press1996}. One of the simplest methods is called the trapezoidal rule because it approximates the area under the curve by a trapezoid,
\begin{flalign}
&\int_{x_1}^{x_N} f(x)dx=\Delta x\left[\frac{1}{2}f_1+f_2+\cdots+f_{N-1}+\frac{1}{2}f_N\right] &\nonumber\\
&+\mathcal{O}\left(\frac{(x_N-x_1)^3 f''}{N^2}\right). &
\end{flalign}
A more sophisticated method is called Simpson's rule, which considers a quadratic interpolation between the points,
\begin{flalign}
& \int_{x_1}^{x_N} f(x)dx=\Delta x\left[\frac{1}{3}f_1+\frac{4}{3}f_2+\frac{2}{3}f_3+\frac{4}{3}f_4+\cdots\right. &\nonumber\\
&\left. +\frac{2}{3}f_{N-2}+\frac{4}{3}f_{N-1}+\frac{1}{3}f_N\right]+\mathcal{O}\left(\frac{(x_N-x_1)^5f^{(4)}}{N^4} \right) . &
\end{flalign}
Either method will successfully calculate the integral necessary for the effective range.

\section{Examples}
\label{sec:examples}

We chose four potentials to illustrate the numerical procedure described in Sec.~\ref{sec:numerical-procedure}. The reasoning behind these choices is the following. The first one is the spherical well. Since we derived analytical expressions for its scattering length and effective range, Eqs.~(\ref{eq:well-scattering_length}) and~(\ref{eq:well-effective-range}), this represents an excellent opportunity to test our program. We can readily compare our numerical solutions to the expected results, and it is much more straightforward to check the correctness of the program.

However, as we stated before, numerical computations would be unnecessary if all potentials had analytical expressions for their low-energy parameters. Hence, next, we considered the modified P\"oschl-Teller potential. In this case, there is an analytical expression for the scattering length, but the effective range has a closed form only when $|a|\to\infty$. Next, we considered a Gaussian potential since, in this situation, everything has to be computed numerically.

In the last three examples, we only considered purely attractive potentials. However, many interesting and relevant physical situations involve potentials where there is competition between a strong short-range repulsion and a weak long-range attraction. For this reason, we also considered the Lennard-Jones potential.

\subsection{Spherical well}

\begin{figure}[!htb] 
\centering
\includegraphics[width=0.5\textwidth]{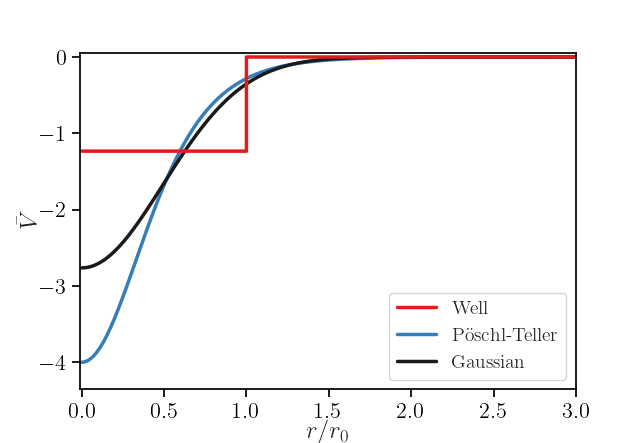}
\caption{Spherical well, modified P\"oschl-Teller, and Gaussian potentials in the unitary limit ($|a|\to \infty$). We make the axis dimensionless by rescaling the distance in effective range units, and the potential in $\hbar^2/(m_r r_0^2)$ units, such that $\bar{V}\equiv m_r r_0^2V/\hbar^2$ is dimensionless.}
\label{fig:unitarity-pot-shape}
\end{figure}

This potential has been covered extensively in the previous sections. To make comparisons easier with the other potentials, we rewrite Eq.~(\ref{eq:square_well_potential}) as
\begin{equation}
V_{\rm sw}(r) = 
\begin{dcases}
-v_{\rm sw}\,\frac{\hbar^2\mu_{\rm sw}^2}{m_r}, \quad &\text{for } r<R,\\
0, \quad & \text{for } r>R,
\end{dcases}
\end{equation}
where we introduced the quantity $\mu_{\rm sw}=1/R$. In Fig.~\ref{fig:unitarity-pot-shape}, we compare its shape with other purely attractive potentials.

\subsection{Modified P\"oschl-Teller}
\label{sec:mPT}

The modified P\"oschl-Teller potential has been used to describe systems in nuclear~\cite{Gezerlis2008} and atomic physics~\cite{Carlson2003}. We present it in the form
\begin{equation}\label{eq:mPT}
V_{\rm PT}(r) =-v_{\rm PT}\frac{\hbar^2}{m_r}\frac{\mu_{\rm PT}^2}{\cosh^2(\mu_{\rm PT} r)}.
\end{equation}
Figure~\ref{fig:unitarity-pot-shape} illustrates its shape. This potential contains two parameters, $v_{\rm PT}$ and $\mu_{\rm PT}$, related to its depth and range. However, associating a range of $R$ to this potential is not as immediate as for the spherical well case. Fortunately, we only need a point $R$ where $V(R) \approx 0$. So, our numerical approach is to look for a value of $R$ such that the potential is negligible $|V(R)| \leqslant \varepsilon$. Using dimensionless quantities, we found that for the mPT, Gaussian, and LJ potentials, $\varepsilon=10^{-15}$ produces the desired results.

The eigenfunctions of the mPT potential may be obtained analytically~\cite{Flugge}, but their derivation is beyond the scope of this work. However, if we perform the substitution $v_{\rm PT} = \lambda(\lambda-1)/2$, there is an analytical expression for the scattering length in terms of the parameters of the potential~\cite{Madeira2019},
\begin{equation}\label{eq:poschl-teller-analytical}
a\mu_{\rm PT} = \frac{\pi}{2}\cot\left(\frac{\pi\lambda}{2}\right) + \gamma + \Psi(\lambda),
\end{equation}
where $\gamma$ is the Euler-Mascheroni constant~\cite{Abramowitz1965} and $\Psi$ is the digamma function~\cite{Abramowitz1965}.

The $|a|\to\infty$ case corresponds to $\lambda=2$ ($\cot(\pi)$ diverges) or $\lambda=-1$ ($\Psi(-1)$ diverges), that is, $v_\text{PT}=1$. For this particular case, the $s$-wave zero-energy radial function takes a relatively simple form~\cite{Landau},
\begin{equation}
\label{eq:ur_mPT_unitarity}
u_0(r) = \frac{\tanh(\mu_{\text{PT}}\,r)}{\tanh(\mu_{\text{PT}}\, R)}.
\end{equation}
Taking its second derivative and substituting it in the radial equation confirms it is the appropriate solution. Then, we can readily calculate the effective range by integrating Eq.~(\ref{eq:effective-range}). In this case, $g_0(r) = 1-r/a=1$, so that
\begin{flalign}
&r_0 = 2 \int_0^R dr\left[1-\frac{\tanh^2(\mu_\text{PT} r)}{\tanh^2(\mu_\text{PT} R)}\right] = &\nonumber \\ 
&2\left[R - \frac{R}{\tanh^2(\mu_\text{PT} R)} + \frac{1}{\mu_\text{PT}\tanh(\mu_\text{PT} R)}\right].&
\end{flalign}
Since $1/\mu_{\rm PT} \sim R$ and the $\tanh(x)$ function converges rapidly to 1 as we increase $x$, we may set $\tanh(\mu R) = 1$. Thus we have that $r_0=2/\mu_{\text{PT}}$ for $v_{\text{PT}} = 1$ (unitarity).

\subsection{Gaussian}

Many interactions in physical systems are modeled through Gaussians, and their convenience and usefulness cannot be overstated. We write the Gaussian potential as
\begin{equation}
V_{\rm g}(r) = -v_{\rm g}\frac{\hbar^2}{m_r}\mu_{\rm g}^2 e^{-r^2\mu_{\rm g}^2}.
\end{equation}
The comparison of this potential with the previous two is illustrated in Fig.~\ref{fig:unitarity-pot-shape}. Although there are no analytical expressions for the scattering length and effective range in this case, numerical investigations in the literature provide benchmarks~\cite{Jeszenszki2018}.

\subsection{Lennard-Jones}

\begin{figure}[!htb] 
\centering
\includegraphics[width=0.5\textwidth]{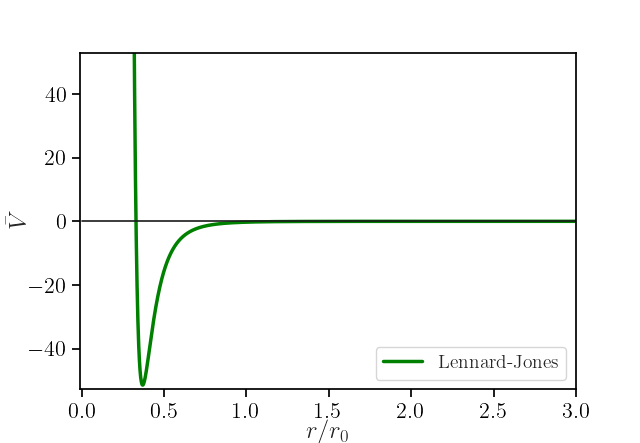}
\caption{
Same as Fig.~\ref{fig:unitarity-pot-shape}, but for the Lennard-Jones potential. Differently from the other potentials, we observe a strong repulsive component near the origin.}
\label{fig:unitarity-pot-shape_LJ}
\end{figure}

To include in our analysis a potential that also contains repulsion, we considered the Lennard-Jones potential,
\begin{equation}
    V_{\rm LJ}(r) =  \frac{\hbar^2}{m_r}\left[\frac{C_{12}}{r^{12}} - \frac{C_6}{r^6}\right],
\end{equation}
where $C_{12}$ and $C_6$ are constants responsible for the strength of the repulsive and attractive interactions, with units of [length]$^{10}$ and [length]$^{4}$, respectively. In Fig.~\ref{fig:unitarity-pot-shape_LJ} we plot this potential for $|a|\to\infty$. The reader may be more familiarized with an alternative way of writing this potential,
\begin{equation}
    V_{\rm LJ}(r) =  4\varepsilon_{\rm LJ} \left[ \left(\frac{\sigma_{\rm LJ}}{r}\right)^{12} - \left(\frac{\sigma_{\rm LJ}}{r}\right)^6 \right],
\end{equation}
but the conversion between the different constants is straightforward.

Besides the repulsive part, the LJ potential differs from the others considered so far because it diverges for $r\to 0$. Physically, the hard core near the origin prevents two particles from coming close together. This poses a numerical difficulty since $V_{\rm LJ}(0)$ diverges. The boundary condition $u(0)=0$ certainly helps, but the potential at the neighboring site, $V_{\rm LJ}(\Delta r)$, is still large and may lead to numerical instabilities. Hence, the first step of the algorithm presented in Sec.~\ref{sec:central_diff} must be slightly changed to circumvent this problem. Since the potential is very large and positive near $r\approx 0$, it prevents the particles from coming close together, i.e., $u\approx 0$ in this region. For this reason, we define a range $0\leqslant r < r_{\text{min}}$ where $u(r) = 0$ and start our integration at $r = r_{\text{min}}$. After this is done, we may proceed as outlined in Sec.~\ref{sec:central_diff}. Using dimensionless units, we found that taking $r_\text{min}$ such that $V(r_\text{min}) \approx 10^{10}$ produces the desired results. 

\subsection{Tuning the parameters}
\label{sec:tuning}

The four potentials we presented have two parameters. In the case of the purely attractive ones, one parameter is associated with the depth of the potential ($v_{\rm sw}$, $v_{\rm PT}$, and $v_{\rm g}$) and another with its range ($\mu_{\rm sw}$, $\mu_{\rm PT}$, and $\mu_{\rm g}$). In contrast, for the LJ potential, they control the attractive ($C_6$) and repulsive ($C_{12}$) components. The numerical procedure we described so far works if the two parameters are known. For example, they could be listed in the literature. However, the situation is often different: the scattering length and effective range are known, and we want to tune the parameters of a particular potential to reproduce the desired $a$ and $r_0$ values. Since we want to match two values and have two free parameters, the correspondence is one-to-one (with the restriction of how many bound states we want).

The Sec.~\ref{sec:central_diff} algorithm can still be used in this case, but we need to apply it repeatedly as we vary the two parameters. Let us denote the values of the two parameters by $v_1$ and $v_2$. The first step is to start with a guess for $v_1$ and $v_2$ and then perform the procedure as described in Sec.~\ref{sec:central_diff} to obtain the scattering length and effective range. They will probably not be the desired values, but it is important to check the radial wave function to see if at least the number of bound states is what we expect. Typically, we want no bound states for $a<0$ and one for $a>0$, in other words, a nodeless $u(r)$ for the first and one node for the latter, as depicted in Fig.~\ref{fig:scattering_intercept}. If the initial guess contains more bound states than desired, then we need to decrease the potential depth until we reach the target number. Then, a possible procedure is:
\begin{enumerate}
\item Start with a guess $(v_1,v_2)$.
\item Compute $a$ and $r_0$ as in Sec.~\ref{sec:central_diff}.
\item Keep $v_2$ fixed. Vary $v_1$ until $a$ has the desired value. Increasing the depth of the potential will decrease the value of the scattering length (until it diverges and changes from $-\infty$ to $+\infty$).
\item Keep $v_1$ fixed at the value found in step 3. Vary $v_2$ until $r_0$ has the desired value. Increasing the range of the potential will increase $r_0$.
\item If $a$ and $r_0$ match the desired values, stop. Else, go to step 3 and use the value of $v_2$ found in step 4.
\end{enumerate}
We may implement this algorithm using two nested loops: one varies $v_1$, and the other runs $v_2$. It is helpful to define a tolerance for the numerical values of $a^{\rm num}$ and $r_0^{\rm num}$, such that $|a^{\rm num}-a|$ and $|r_0^{\rm num}-r_0|$ are small quantities. After the procedure is done, checking the number of nodes of the radial function is necessary to guarantee that it is indeed the situation we wanted to reproduce.

\subsection{Results}

Finally, applying the numerical procedure described so far to some concrete examples is instructive. The objective is twofold: we want to illustrate the method and provide some results so the reader can use them to check their program. We present 3 cases: $a<0$, $|a|\to\infty$, and $a>0$, which correspond to three very distinct physical situations.

As we pointed out in Sec.~\ref{sec:approximations}, the universality in the low-energy scattering results is not specific to a single physics area so we can choose from several instances. We picked nuclear physics since two nucleon systems are excellent for illustrating low-energy scattering. There is no bound state in the region where $a$ is negative. That is the case of two neutrons, with a scattering length of $a=-18.5$ fm and an effective range of $r_0 = 2.7$ fm~\cite{Gezerlis2008}. The region where $a>0$ is interesting due to the appearance of the first bound state of the system, where lies the deuteron, a proton and a neutron, with $a = 5.4$ fm and $r_0 = 1.7$ fm. The region where $|a|\to \infty$, often called unitary limit, is of great interest~\cite{Randeria2014}. It does not occur naturally in nuclear physics, but we can still study its properties. Table~\ref{tab:examples} summarizes the physical situations we cover in this section.

\renewcommand{\arraystretch}{1.05}
\begin{table}[!htb]
    \centering
    \caption{Scattering length and effective range values that we want to reproduce with four forms of two-body potentials.}
    \begin{tabular}{ccc}
    \hline\hline
    System   & $a$ (fm) & $r_0$ (fm) \\
    \hline
    \textbf{Neutron-neutron} & $-18.5$ & $2.7$  \\
    \textbf{Unitarity} & $\pm \infty$ & $1.0$ \\
    \textbf{Deuteron} &  $5.4$ & $1.7$ \\
    \hline\hline
    \end{tabular}
    \label{tab:examples}
\end{table}

As a side note, it is worth mentioning that most systems have their interactions fixed by what is observed in nature, such as the case of two neutrons, the deuteron, and countless other examples. Although it is possible to theoretically investigate the continuous variation of the scattering length from negative to positive values, we would like to see this happening in a physical system. In experiments with cold atoms, this is possible. A clever method called Feshbach resonances~\cite{Moerdijk1995} uses the manipulation of magnetic fields to vary the scattering length of the potential between two atoms. This allows the study of systems with scattering lengths of both signs and the unitarity limit between them.

We performed the procedure described in Sec.~\ref{sec:tuning} to obtain the parameters, scattering lengths, and effective ranges presented in Tables~\ref{tab:parameters_3pot} and~\ref{tab:LJ_table}. It is worth remembering that we can check the results for the spherical well against the analytical expressions, Eqs.~(\ref{eq:well-scattering_length}) and~(\ref{eq:well-effective-range}). For the mPT, Eq.~(\ref{eq:poschl-teller-analytical}) provides a benchmark for the scattering length, and in Sec.~\ref{sec:mPT} we show that $r_0=2/\mu_{\rm PT}$ when $|a|\to\infty$.

\renewcommand{\arraystretch}{1.2}
\begin{table}[!htb]
    \centering
    \caption{Parameters for the spherical well, modified P\"oschl-Teller, and Gaussian potentials that reproduce the desired scattering lengths and effective range.}
    \begin{tabular}{ccccc}
    \hline\hline
    Potential   & $v$ & $\mu$ (fm$^{-1}$)& $a$ (fm) & $r_0$ (fm) \\
    \hline
    \multicolumn{5}{c}{\textbf{Neutron-neutron}}\\
    \hline
    Well & $1.1096$ & $0.3918$ & $-18.52$ & $2.7$ \\
    mPT & $0.9071$ & $0.7991$ & $-18.51$& $2.7$  \\
    Gaussian & $1.2121$ & $0.5672$ & $-18.55$ & $2.7$ \\
    \hline
    \multicolumn{5}{c}{\textbf{Unitarity}} \\ \hline
    Well & $1.2337$ & $1.0000$ & $\sim -10^5$ & $1.0$ \\
    mPT & $1.0000$ & $2.0000$ & $\sim 10^{9}$& $1.0$  \\
    Gaussian & $1.3420$ & $1.4349$ & $\sim -10^{5}$ & $1.0$ \\
    \hline
    \multicolumn{5}{c}{\textbf{Deuteron}}\\
    \hline    
    Well & $1.7575$ & $0.5000$ & $5.4$ & $1.70$ \\
    mPT & $1.4388$ & $0.8631$ & $5.4$ & $1.73$  \\
    Gaussian & $1.9102$ & $0.6754$ & $5.4$ & $1.70$ \\
    \hline\hline
    \end{tabular}
    \label{tab:parameters_3pot}
\end{table}

\begin{table}[!htb]
    \centering
    \caption{Numerical results for the Lennard-Jones potential.}
    \begin{tabular}{ccccc}
    \hline\hline
    System   & $C_{12}$ (fm$^{10}$)& $C_6$ (fm$^{4}$)& $a$ (fm) & $r_0$ (fm) \\
    \hline
    \textbf{Neutron-neutron} & $3.08836698$ & $9.86668911$ & $-18.5$& $2.71$  \\
    \textbf{Unitarity} & $0.00034068$ & $0.26462461$ & $\sim -10^{5}$ & $1.00$ \\
    \textbf{Deuteron} & $0.90485319$ & $6.81472000$ & $5.4$ & $1.70$ \\
    \hline\hline
    \end{tabular}
    \label{tab:LJ_table}
\end{table}

The two-body potentials are very different in shape, as shown in Figs.~\ref{fig:unitarity-pot-shape} and~\ref{fig:unitarity-pot-shape_LJ}. The fact that we can tune them to produce the same scattering length $a$ and effective range $r_0$, although remarkable, is directly related to the shape-independent approximation, Eq.~(\ref{eq:shape_independent_equation}). As the name suggests, this equation describes the $s$-wave phase shift $\delta_0(k)$ as independent of the shape of the scattering potential.

\begin{figure}[!htb] 
    \centering
    \begin{minipage}{0.5\textwidth}
        \centering
        \subfloat[Neutron-neutron: $a= -18.5$ fm and $r_0 = 2.7$ fm. \label{fig:radial_solution_nn}]{\includegraphics[width=0.95\textwidth]{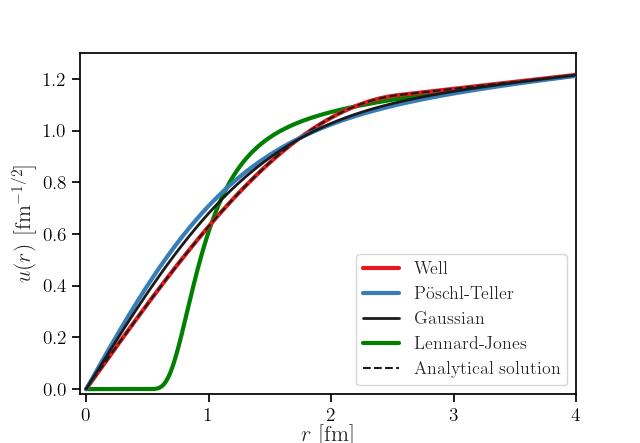}}
    \end{minipage}\hfill
    \begin{minipage}{0.5\textwidth}
        \centering
        \subfloat[Unitarity: $|a|\to \infty$ and $r_0 = 1.0$ fm. \label{fig:radial_solution_uni}]{\includegraphics[width=0.95\textwidth]{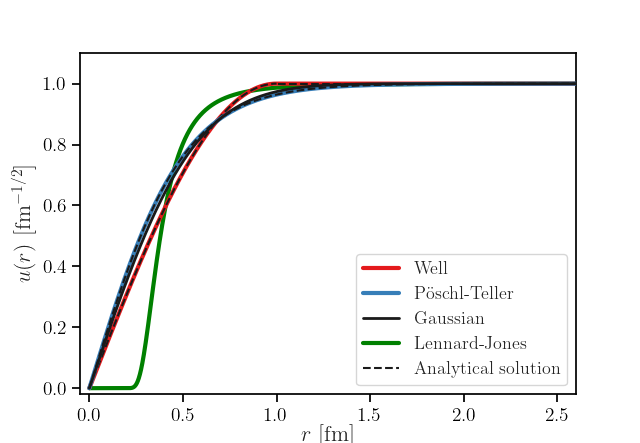}}
    \end{minipage}\hfill
    \begin{minipage}{0.5\textwidth}
        \centering
        \subfloat[Deuteron: $a=5.4$ fm and $r_0 = 1.7$ fm. \label{fig:radial_solution_deut}]{\includegraphics[width=0.95\textwidth]{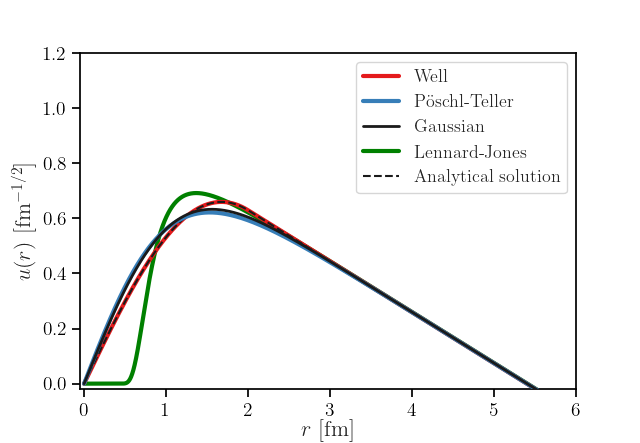}}
    \end{minipage}
\caption{Reduced radial solutions, $u(r) = rA(r)$, of Schrödinger's equation with the proposed potentials for three different situations. The dashed curves correspond to the analytical solutions. Outside the range of the potential, large $r$, all solutions must be the same.}
    \label{fig:radial_solution}
\end{figure}

Besides the values of $a$ and $r_0$, the numerical procedure produces the reduced radial wave functions $u(r)$. In Fig.~\ref{fig:radial_solution}, we compare $u(r)$ solved with the Numerov's method for the spherical well, modified P\"oschl-Teller, Gaussian, and Lennard-Jones potentials. They are in excellent agreement with analytical solutions, Eq.~(\ref{eq:well_k_lim_scat_solutions}) for the spherical well, and Eq.~(\ref{eq:ur_mPT_unitarity}) for the mPT at unitarity. For the same physical system, the functions differ inside the range of the potentials but are all the same outside, $1-r/a$.

It is also possible to see the geometric interpretation of the scattering length alluded to in Figs.~\ref{fig:fig:scattering_intercept_a} and~\ref{fig:scattering_intercept_b}. For $a<0$, Fig.~\ref{fig:radial_solution_nn}, $u(r)$ does not intercept the $r$-axis, and the scattering length is negative because the extrapolation of the reduced radial function intercepts the axis at a negative value. In Fig.~\ref{fig:radial_solution_deut}, $u(r)$ intercepts the $r$-axis at the scattering length, thus $a>0$. Neither the reduced radial function nor its extrapolation intercept the axis at unitarity, since $|a|\to\infty$, Fig.~\ref{fig:radial_solution_uni}.

\begin{figure}[!htb] 
    \centering
    \begin{minipage}{0.5\textwidth}
        \centering
        \subfloat[Spherical well, modified P\"oschl-Teller, and Gaussian potentials. \label{fig:BCS-BEC_model-potentials}]{\includegraphics[width=\textwidth]{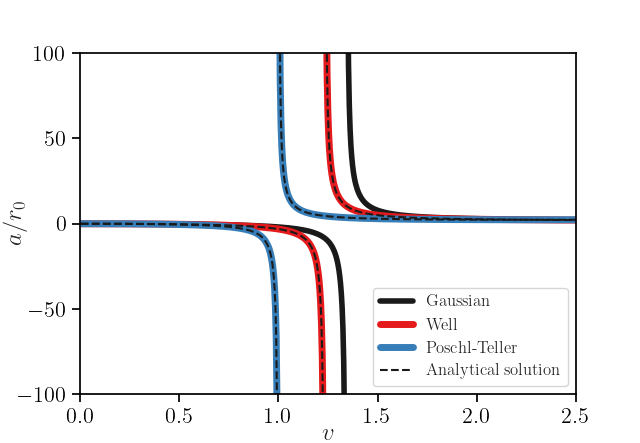}}
    \end{minipage}
    \begin{minipage}{0.5\textwidth}
        \centering
        \subfloat[Lennard-Jones potential. \label{fig:LJ_a}]{\includegraphics[width=\textwidth]{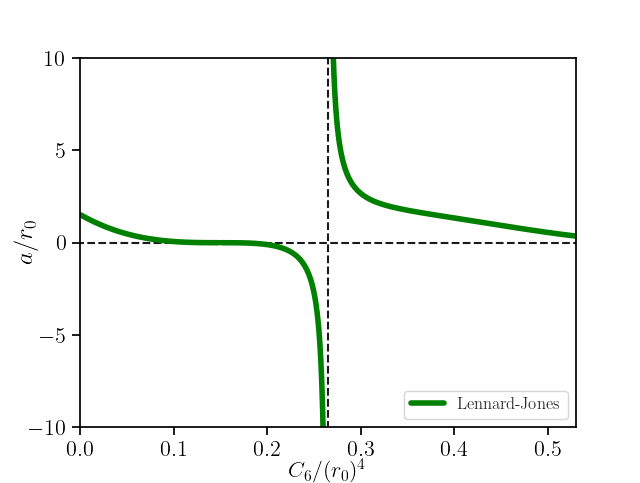}}
    \end{minipage}\hfill
        \caption{Scattering length behavior as a function of the interaction strength for the spherical well, modified P\"oschl-Teller, Gaussian, and Lennard-Jones potentials. The dashed curves in (a) are the analytical solutions for the spherical well, Eq.~(\ref{eq:well-scattering_length}), and for the modified P\"oschl-Teller potential, Eq.~(\ref{eq:poschl-teller-analytical}).}
    \label{fig:BCS-BEC}
\end{figure}

Finally, after obtaining the parameters for these three illustrative physical systems, it is interesting to investigate how the scattering length varies as we continuously vary the parameters of the potentials. In Fig.~\ref{fig:BCS-BEC_model-potentials}, we present the ratio $a/r_0$ as a function of the potential depth of the spherical well, modified P\"oschl-Teller, and Gaussian potentials ($v_{\rm sw}$, $v_{\rm PT}$, and $v_{\rm g}$). Analytical expressions are only available for the first two, and it is possible to see that our numerical investigations are in excellent agreement with them.

To produce a similar figure for the Lennard-Jones case, we fixed the value of $C_{12}/r_0^{10}=0.000341$ and varied $C_6$. We present the results in Fig.~\ref{fig:LJ_a}. Although the behavior is qualitatively similar to the purely attractive potentials, the main difference is the region close to $C_6\approx 0$ where the scattering length is positive. This is because, for $C_6=0$, the Lennard-Jones potential is purely repulsive. Then the scattering length is positive not due to a bound state but because the repulsive part ``pushes'' the wave function away from the origin, as depicted in Fig.~\ref{fig:scattering_intercept_c}.

\section{Conclusions}
\label{sec:conclusion}

This work presented quantum scattering theory fundamentals focusing on the low-energy limit. In this context, we aimed to introduce two significant quantities: the scattering length and the effective range. To illustrate how these two parameters behave in a concrete example, we derived analytical expressions for both in the case of the spherical well. We also showed how the energy of a bound state could be calculated using zero- and finite-range expressions applied to a $^4$He dimer and the deuteron.

After introducing the main results of low-energy scattering theory, we proceeded with our main goal: describing a numerical procedure that can be used to compute the scattering length and effective range of any spherically symmetric finite-ranged two-body potential. We chose the spherical well, modified P\"oschl-Teller, Gaussian, and Lennard-Jones potentials as examples. We then performed calculations related to three physical systems: two neutrons, unitarity, and the deuteron.

This work shows that these calculations are relevant and can be readily applied in atomic and nuclear physics. Hopefully, students can follow the procedure outlined in this manuscript, reproduce our results, extend them to their choice of physical systems, and apply the method to other potentials.

\begin{acknowledgments}
This work was partially supported by the S\~ao Paulo Research Foundation (FAPESP) grant 2018/09191-7 (L.M.), and by the National Council for Scientific and Technological Development (CNPq) grants 383501/2022-9 (L.M.) and 131025/2022-8 (M.M.L.).
\end{acknowledgments}

\appendix


\section{Scattering theory}
\label{sec:scattering_app}

\subsection{Time-dependent formalism}

We may treat interactions more simply in a formalism called interaction picture. We  denote the time evolution of a state ket at an initial time $t_0$ in the Schrödinger picture as
\begin{equation}\label{eq:int_pic_evolution}
\ket{\phi(t)}_S = U_S(t,t_0)\ket{\phi(t_0)}_S,
\end{equation}
where $|\,\,\rangle_S$ is a ket in the Schrödinger picture and $U_S(t,t_0)$ is the time-evolution operator. If $H$ is time independent, then $U_S(t,t_0) = e^{-iH(t-t_0)/\hbar}$. The interaction-picture state ket, $|\,\,\rangle_I$, is defined as
\begin{equation}
\ket{\phi(t)}_I = e^{iH_0t/\hbar}\ket{\phi(t)}_S.
\end{equation}
The operators are defined as
\begin{equation}
A_I = e^{iH_0t/\hbar}A_Se^{-iH_0t/\hbar}.
\end{equation}

The Schrödinger equation $i\hbar\frac{\partial}{\partial t} \ket{\phi(t)}_S = H\ket{\phi(t)}_S$ may be written as
\begin{equation}\label{eq:interaction_pic_schrodinger}
i\hbar\frac{\partial}{\partial t} \ket{\phi(t)}_I = V_I\ket{\phi(t)}_I,
\end{equation}
where $V_I = e^{iH_0(t-t_0)/\hbar}Ve^{-iH_0(t-t_0)/\hbar}$ is the potential in the interaction picture. Equation~(\ref{eq:interaction_pic_schrodinger}) is a Schrödinger-like equation. The advantage is that we remove $H_0$ from our calculations to consider the interaction. If $V=0$, $\ket{\phi(t)}_I$ is constant in time (and equal to $\ket{\phi(t_0)}_S$). Similar to Eq.~\ref{eq:int_pic_evolution}, $\ket{\phi(t)}_I$ evolves in time as
\begin{equation}\label{eq:interaction_pic-time-evo}
\ket{\phi(t)}_I = U_I(t,t_0)\ket{\phi(t_0)}_I, 
\end{equation}
where $U_I(t,t_0)$ is defined as 
\begin{equation}\label{eq:U_I-definition}
U_I(t,t_0) = e^{iH_0t/\hbar}U_S(t,t_0)e^{-iH_0t_0/\hbar}
\end{equation}
and obeys the Schrödinger-like equation
\begin{equation}\label{eq:U_I-equation}
i\hbar\frac{\partial}{\partial t} U_I(t,t_0) = V_I(t)U_I(t,t_0).
\end{equation}

The solution of Eq.~\ref{eq:U_I-equation} may be formally written as
\begin{equation}\label{eq:dirac_picture-time_evolution-solution}
U_I(t,t_0) = 1 - \frac{i}{\hbar}\int_{t_0}^t dt' V_I(t')U_I(t',t_0),
\end{equation}
for the required initial condition $U_I(t_0,t_0) = 1$. Our goal is to calculate the evolution of the state in a distant past $t_0\to -\infty$, that is, $\ket{\phi (t \to -\infty)} = \ket{i}$. Nevertheless, equation~(\ref{eq:dirac_picture-time_evolution-solution}) is only valid for finite times $t$ and $t_0$ and so setting $U_I(t,-\infty)$ would lead to convergence problems. To avoid it, we will give meaning to $U_I(t,-\infty)$ by writing it as~{\cite{Gellmann1953}} (proof in Appendix~\ref{appendix:A}),
\begin{equation}
U_I(t,-\infty) = \lim_{t_0\to -\infty} U_I(t,t_0) = \lim_{\epsilon \to 0} \epsilon \int_{-\infty}^{0} dt'\,e^{\epsilon t'}U_I(t,t'),
\end{equation}
where $U_I(t,t')$ is defined as in eq.~(\ref{eq:U_I-definition}). Despite our time-dependent treatment, we recall that $H$ is time-independent. Thus $U_S(t,t') = e^{-iH(t-t')/\hbar}$. The state vector at a time $t= 0$ is given by
\begin{equation}\label{eq:state_at_t=0}
\ket{\phi(t=0)}_I = U_I(0,-\infty)\ket{i},
\end{equation}
where
\begin{equation}
U_I(0,-\infty) = \lim_{\epsilon \to 0} \epsilon \int_{-\infty}^{0} dt'\,e^{\epsilon t'}e^{iHt'/\hbar}e^{-iH_0t'/\hbar}.
\end{equation}

Now applying it to equation~(\ref{eq:state_at_t=0}):
\begin{flalign}\label{eq:time_evolution_integral_operator}
&\ket{\phi(t=0)}_I = \lim_{\epsilon \to 0} \epsilon \int_{-\infty}^{0} dt'\,e^{\epsilon t'}e^{i(H-E_i)t'/\hbar}\ket{i} = &\nonumber \\
&\lim_{\epsilon\to 0} \frac{i\epsilon}{E_i - H +i\epsilon}\ket{i}.&
\end{flalign}
Using the identity (proof in Appendix~\ref{appendix:B})
\begin{flalign}\label{eq:resolvent-identity}
&\frac{1}{E_i-H+i\epsilon} - \frac{1}{E_i-H_0+i\epsilon} = &\nonumber \\
&\frac{1}{E_i-H_0+i\epsilon}V\frac{1}{E_i-H+i\epsilon},&
\end{flalign}
we rewrite Eq.~(\ref{eq:time_evolution_integral_operator}) as
\begin{align}
&\ket{\phi(t=0)}_I =\lim_{\epsilon\to 0} \frac{i\epsilon}{E_i - H_0 +i\epsilon}\ket{i} + \nonumber \\
&\frac{1}{E_i - H_0 +i\epsilon}V\frac{i\epsilon}{E_i - H +i\epsilon}\ket{i} = \nonumber \\ 
&\lim_{\epsilon\to 0} \frac{i\epsilon}{E_i - H_0 +i\epsilon}\ket{i} + \frac{1}{E_i - H_0 +i\epsilon}V\ket{\phi(t=0)}_I
\end{align}
Since $\ket{i}$ is an eigenfunction of $H_0$ (it satisfies $H_0\ket{i} = E_i\ket{i}$), the first term in the RHS is the identity operator. Then we can write
\begin{equation}\label{eq:lippmann-schwinger}
\ket{\psi} = \ket{i} + \frac{1}{E_i-H_0+i\epsilon}V\ket{\psi},
\end{equation}
where we left off the notation $\ket{\phi(t=0)}$ to emphasize that this is an actual time-independent problem. This is know as the Lippmann-Schwinger equation.

Back to the time-dependent formulation, the state vector at any time $t$ could be obtained by the sequential relation of the time translation operator: $U_I(t,t_0) = U_I(t,t')U_I(t',t_0)$, so
\begin{equation}
\ket{\phi(t)} = U_I(t,0)\ket{\phi(0)} = U_I(t,-\infty)\ket{i}.
\end{equation}
In a distant future, that is, setting $t\to +\infty$, we write
\begin{equation}\label{eq:S_matrix-evolution}
\ket{f} = S\ket{i},
\end{equation}
where $\ket{\phi(t\to+\infty)}$ is our asymptotic final state $\ket{f}$ and the scattering matrix (or $S$ matrix) is defined as
\begin{equation}
S \equiv U_I(+\infty,-\infty).
\end{equation}
Equation~(\ref{eq:S_matrix-evolution}) asserts that the action of the $S$ matrix on an initial state (that exists asymptotically for $t_0\to -\infty$) transforms the ket $\ket{i}$ into a final state that exists in a distant future $t\to +\infty$. That is, some sort of scattering occurred between $t$ and $t_0$. Furthermore, Eq.~(\ref{eq:lippmann-schwinger}) may be written as the power series expansion
\begin{align*}
\ket{\psi} =& \ket{i} + G_+V\ket{i} + G_+VG_+V\ket{i}+...\\
&=\ket{i} + G_+\left(V + VG_+V+... \right)\ket{i},
\end{align*}
where $G_+\equiv(E_i-H_0+i\epsilon)^{-1}$.

We define the transition matrix $T$ as the perturbative series
\begin{equation}\label{eq:T-matrix}
T \equiv V + VG_+V + VG_+VG_+V+...,
\end{equation}
which is often called the Dyson series. We may write the scattering state as
\begin{equation}
\ket{\psi}  = \ket{i} + G_+T\ket{i}.
\end{equation}
The consequence is that
\begin{equation}
V\ket{\psi} = T\ket{i}.
\end{equation}
The perturbative character of Eq.~(\ref{eq:T-matrix}) relates the $T$-matrix to being a species of a generalized potential, where in a first-order perturbation $T$ and $V$ are equivalent and thus $\ket{\psi} \approx \ket{i}$. This is known as the first-order Born approximation~{\cite{Sakurai2014}}.

\subsection{Scattering theory integral equations}

Now we focus on the physical implications of Eq.~(\ref{eq:lippmann-schwinger}). Writing $\ket{\psi}$ in the position basis and inserting a complete set of position states between $G_+$ and $V$ leads to the integral equation
\begin{equation}\label{eq:pos_representation_generalpsi}
\braket{\mb{r}}{\psi} = \braket{\mb{r}}{i} + \int d^3\mb{r}' \element{\mb{r}}{G_+}{\mb{r}'}\element{\mb{r}'}{V}{\psi},
\end{equation}
where we must calculate the function
\begin{equation}
G_+(\mb{r},\mb{r}') \equiv \frac{\hbar^2}{2m}\left\langle\mb{r}\left|\frac{1}{E-H_0+i\epsilon}\right|\mb{r}'\right\rangle.
\end{equation}
Recalling that the momentum basis $\{\ket{\mb{k}}\}$ set elements are eigenstates of $H_0$ with eigenvalues $\hbar^2\mb{k}^2/2m$, Eq.~(\ref{eq:free_eigenvalue_eq}), we may insert this discrete basis by writing
\begin{equation}
G_+(\mb{r},\mb{r}') = \frac{\hbar^2}{2m} \sum_{\mb{k}',\mb{k}''} \braket{\mb{r}}{\mb{k}'}\left\langle\mb{k}'\left|\frac{1}{E-H_0+i\epsilon}\right|\mb{k}''\right\rangle \braket{\mb{k}''}{\mb{r}'}.
\end{equation}

The quantized plane waves in the position representation are written as
\begin{equation}
\braket{\mb{r}}{\mb{k}} = \frac{e^{i\mb{k}\cdot\mb{r}}}{L^{3/2}} \,\,\text{ and } \braket{\mb{k}}{\mb{r}} = \frac{e^{-i\mb{k}\cdot\mb{r}}}{L^{3/2}}.
\end{equation}
Now by letting $H_0$ act on $\bra{\mb{k}'}$ and noting that $\braket{\mb{k}'}{\mb{k}''} = \delta_{\mb{k}'\mb{k}''}$ and $E = \hbar^2k^2/2m$,
\begin{equation}
G_+(\mb{r},\mb{r}') = \frac{1}{L^3} \sum_{\mb{k}'} \frac{e^{i\mb{k}'\cdot(\mb{r}-\mb{r}')}}{k^2-k'^2+i\epsilon},
\end{equation}
where we absorbed the factor $2m/\hbar^2$ into $\epsilon$.

We are left with a sum in $k$-space, a discrete space. Due to periodic boundary conditions, each $k_i$ ($i =x,y,z$) is located at $k_i = 2\pi n_i/L$, where $n_i = 0,1,2,3...$. That is, $\{k\} = \{0,2\pi/L,4\pi/L,...\}$. The distance between each point in the $k$-space is $\Delta k = 2\pi/L$. We must take $L\to \infty$ to guarantee the required continuous character. For this reason, the separation between points is very small compared to $L$ ($\Delta k \approx 0$). The summation in $\mb{k}$ may be taken to be over a continuous space. We may substitute the sum with an integral,
\begin{equation}
\sum_{\mb{k}'} \to  \int \rho(k)d^3\mb{k}' = \frac{L^3}{(2\pi)^3}\int d^3\mb{k}',
\end{equation}
where the factor $\rho(k) = L^3/(2\pi)^3$ is the $k$-density in 3 dimensions. The function becomes
\begin{equation}
G_+(\mb{r},\mb{r}') = \frac{1}{(2\pi)^3}\int d^3\mb{k}'\frac{e^{i\mb{k}'\cdot(\mb{r}-\mb{r}')}}{k^2-k'^2+i\epsilon}.
\end{equation}
We may perform this integration by choosing spherical coordinates $(k', \theta, \phi)$ and, without loss of generalization, we may let the vector $(\mb{r}-\mb{r}')$ lie along the $k'_z$ axis so that $\mb{k}'\cdot(\mb{r}-\mb{r}') = k'|\mb{r}-\mb{r}'|\cos\theta$. We then write
\begin{align}
G_+(\mb{r},\mb{r}') &= \frac{1}{(2\pi)^2}\int_0^{\infty} dk' k'^2  \int_0^\pi d\theta \frac{e^{ik'|\mb{r}-\mb{r}'|\cos\theta}}{k^2-k'^2+i\epsilon} \nonumber\\
&= \frac{1}{4\pi^2|\mb{r}-\mb{r}'|}\int_{-\infty}^{\infty}dk' \frac{k'\sin(k'|\mb{r}-\mb{r}'|)}{k^2-k'^2+i\epsilon},
\end{align}
where we note that the integrand is even. We also note a pole in $k' = \pm \sqrt{k^2+i\epsilon}$. It would make it ambiguous if we chose to take $\epsilon \to 0$ before evaluating the integral. Luckily, the pole is not on the real axis as $\pm\sqrt{k^2+i\epsilon} \approx k +i\epsilon/2k + \epsilon^2/8k^3 + ...$, which gives meaning to the integration process. Ignoring terms higher than $\epsilon^2$ and redefining $\epsilon/2k \to \epsilon$, we may factorize $(k^2-k'^2+i\epsilon) = -(k'-k-i\epsilon)(k'+k+i\epsilon)$. The integral is then
\begin{equation}\label{eq:G_integral}
G_+(\mb{r},\mb{r}') = \frac{1}{8i\pi^2|\mb{r}-\mb{r}'|}\int_{-\infty}^{\infty}dk' \frac{k'(e^{-ik'|\mb{r}-\mb{r}'|}-e^{ik'|\mb{r}-\mb{r}'|})}{(k'-k-i\epsilon)(k'+k+i\epsilon)}.
\end{equation}

\begin{figure}[!htb] 
    \centering
    \begin{minipage}{0.5\textwidth}
        \centering
        \subfloat[Upper plane path where $e^{ik|\mb{r}-\mb{r}'|}\to 0$. \label{subfig:upper-plane}]{\includegraphics[width=\textwidth]{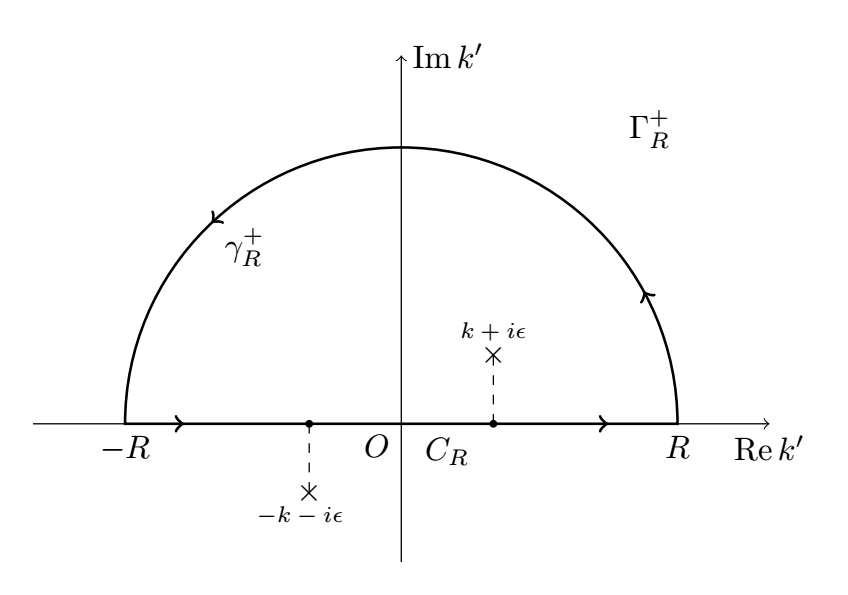}}
    \end{minipage}\hfill
    \begin{minipage}{0.5\textwidth}
        \centering
        \subfloat[Lower plane path where $e^{-ik|\mb{r}-\mb{r}'|}\to 0$. \label{subfig:lower-plane}]{\includegraphics[width=\textwidth]{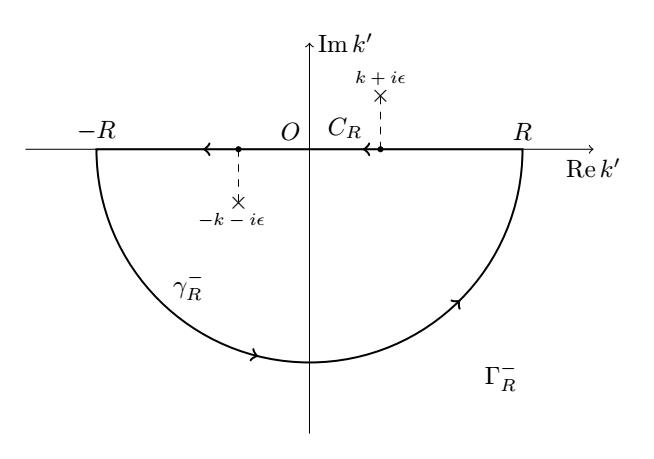}}
    \end{minipage}
        \caption{Paths to calculate the integral~(\ref{eq:G_integral}). We must choose the upper plane path $\Gamma^{+}_R$ for $e^{+ik'|\mb{r}-\mb{r}'|}$ because $e^{+ik'|\mb{r}-\mb{r}'|}\to 0$ as $\operatorname{Im}k'$ takes $+\infty$ values. Similarly, we choose the lower plane path $\Gamma^{-}_R$ for $e^{-ik'|\mb{r}-\mb{r}'|}$ because $e^{-ik'|\mb{r}-\mb{r}'|}\to 0$ as $\operatorname{Im}k'$ takes $-\infty$ values. This makes the integration in $\gamma^{\pm}_R$ go to $0$, and we are left with the path $C_R$, which lies in the axis $\operatorname{Re}k'$. That is, the closed path $\Gamma^{\pm}_R$ is equal to the real integral~(\ref{eq:G_integral}).}
    \label{fig:integral_paths}
\end{figure}

We should let $k'$ momentarily be a complex variable to carry out this integration and use the residue theorem~\cite{Arfken2005}. Consider the paths in Fig.~\ref{fig:integral_paths}. The closed path integral may be written as the sum
\begin{equation}
\oint_{\Gamma_R}  = \int_{\gamma_R} + \int_{C_R} = 2\pi i\times \sum_j\text{Res}\{k';j\},
\end{equation}
where the integral in the closed contour $\Gamma_R$ is $2\pi i$ times the sum of the residues $\text{Res}\{k';j\}$ due to poles $j$ enclosed by the curve, and the integral through the path $\gamma_R$ is zero due to Jordan's lemma. The only pole inside $\Gamma^{\pm}_R$ is $k\pm i\epsilon$. The residues may be calculated as follows
\begin{align}
&\text{Res}\{k';k+i\epsilon\} = \nonumber\\
&\lim\limits_{\substack{%
    {k'\to k+i\epsilon}\\
    \epsilon\to 0}} (k'-k-i\epsilon)\,\,\frac{k'e^{ik'|\mb{r}-\mb{r}'|}}{(k'-k-i\epsilon)(k'+k+i\epsilon)} = \frac{e^{ik|\mb{r}-\mb{r}'|}}{2}, \\[5ex]
&\text{Res}\{k';-k-i\epsilon\} = \nonumber \\
&\lim\limits_{\substack{%
    {k'\to -k-i\epsilon}\\
    \epsilon\to 0}} (k'+k+i\epsilon)\,\,\frac{k'e^{-ik'|\mb{r}-\mb{r}'|}}{(k'-k-i\epsilon)(k'+k+i\epsilon)} = \frac{e^{ik|\mb{r}-\mb{r}'|}}{2}.
\end{align}
The integral through the real axis $C_R$ is
\begin{equation}
\lim_{R\to \infty}\int_{-R}^{R}\,dk' \frac{k'(e^{-ik'|\mb{r}-\mb{r}'|}-e^{ik'|\mb{r}-\mb{r}'|})}{(k'-k-i\epsilon)(k'+k+i\epsilon)}= 2\pi i\times e^{ik|\mb{r}-\mb{r}'|}.
\end{equation}
Thus
\begin{equation}
G_+(\mb{r},\mb{r}') = \frac{1}{4\pi}\frac{e^{ik|\mb{r}-\mb{r}'|}}{|\mb{r}-\mb{r}'|}.
\end{equation}

Returning to Eq.~(\ref{eq:pos_representation_generalpsi}):
\begin{equation}
\braket{\mb{r}}{\psi} = \braket{\mb{r}}{i} - \frac{2m}{\hbar^2}\int d^3\mb{r}' \frac{1}{4\pi}\frac{e^{ik|\mb{r}-\mb{r}'|}}{|\mb{r}-\mb{r}'|}\element{\mb{r}'}{V}{\psi}.
\end{equation}
Considering the potential $V(\mb{r}')$ to be local, that is,
\begin{equation}
\element{\mb{r}'}{V}{\mb{r}''} = V(\mb{r}')\delta(\mb{r}'-\mb{r}'').
\end{equation}
This allows us to write
\begin{equation}
\braket{\mb{r}}{\psi} = \braket{\mb{r}}{i} - \frac{2m}{\hbar^2}\int d^3\mb{r}' \frac{1}{4\pi}\frac{e^{ik|\mb{r}-\mb{r}'|}}{|\mb{r}-\mb{r}'|}V(\mb{r}')\braket{\mb{r}'}{\psi}.
\end{equation}

We now restrict ourselves to cases where the potential is of a finite range. This condition is important because the scattering is observed far away from the scattering center. We are interested in measurements at large distances $|\mb{r}|\gg |\mb{r}'|$. In this regime,
\begin{equation}
e^{ik|\mb{r}-\mb{r}'|} \approx e^{ikr}e^{-i\mb{k}'\cdot\mb{r}'},
\end{equation}
where $r\equiv |\mb{r}|$. Furthermore, we now specify that our initial state is $\ket{i} = \ket{\mb{k}}$ and $\braket{\mb{r}}{\mb{k}} = e^{i\mb{k}\cdot\mb{r}}/L^{3/2}$. Finally, we arrive at the form
\begin{equation}
\psi(\mb{r},\theta) \xrightarrow{\text{large }r} \frac{1}{L^{3/2}}\left[e^{i\mathbf{k}\cdot \mathbf{r}}+\frac{e^{ikr}}{r}f(\mb{k}',\mb{k})\right],
\end{equation}
where 
\begin{equation}
f(\mb{k}',\mb{k}) = -\frac{mL^3}{2\pi\hbar^2}\int d^3x' \braket{\mb{k}'}{\mb{r}'}V(\mb{r}')\braket{\mb{r}'}{\psi}	
\end{equation}
is referred to as the scattering amplitude.

\section{}
\label{appendix:A}

We follow a procedure present in Ref.~\cite{Roman1965}. Consider a well-behaved function $f(t)$ such that
\begin{equation}
\lim_{t_0\to -\infty}f(t_0) = f(-\infty).
\end{equation}
Now consider
\begin{equation}
\lim_{\epsilon\to 0} \epsilon\int_{-\infty}^{0}dt'\,e^{\epsilon t'} f(t'),
\end{equation}
where $\epsilon \ll 1$. Integrating by parts leads to
\begin{align}
&\lim_{\epsilon\to 0}\epsilon\int_{-\infty}^{0}dt'\,e^{\epsilon t'} f(t') = &\nonumber \\
&\left[f(t')e^{\epsilon t'}\right]_{-\infty}^{0} - \lim_{\epsilon\to 0} \int_{-\infty}^{0}dt'\,\frac{df}{dt'}e^{\epsilon t'}& \nonumber\\
&=f(0) - \int_{-\infty}^{0}dt'\,\frac{df}{dt'} = f(-\infty).&
\end{align}
Thus, we have the equality
\begin{equation}
\lim_{t_0\to -\infty}f(t_0) =\lim_{\epsilon\to 0} \epsilon\int_{-\infty}^{0}dt'\,e^{\epsilon t'} f(t').
\end{equation}
We may apply this prescription to the time evolution operator to give a mathematical meaning to $U_I(t,-\infty)$ or $U_I(+\infty,t)$. Thus 
\begin{align}
&U_I(t,-\infty) = \lim_{\epsilon\to 0}\epsilon\int_{-\infty}^{0}dt'\,e^{\epsilon t'} U_I(t,t'),\\
&U_I(+\infty,t_0) = \lim_{\epsilon\to 0}\epsilon\int_{0}^{+\infty}dt'\,e^{-\epsilon t'} U_I(t',t_0).
\end{align}

\section{}
\label{appendix:B}

By defining $G = (E_i-H+i\epsilon)^{-1}$ and $G_0 = (E_i-H_0+i\epsilon)^{-1}$ and noting that $G_0^{-1}-G^{-1} = H-H_0 = V$, we take the difference $G-G_0$ as
\begin{align}
    &G - G_0 = (G_0 G_0^{-1})G - G_0(G^{-1}G) = \nonumber \\
    &G_0(G_0^{-1}-G^{-1})G = G_0VG,
\end{align}
which is the identity in Eq.~(\ref{eq:resolvent-identity}):
\begin{align}
    &\frac{1}{E_i-H+i\epsilon} - \frac{1}{E_i-H_0+i\epsilon} =\nonumber \\
    &\frac{1}{E_i-H_0+i\epsilon}V\frac{1}{E_i-H+i\epsilon}.
\end{align}
Alternatively, we can also write
\begin{align}
    &G - G_0 = G(G_0^{-1}G_0) - (GG^{-1})G_0 = \nonumber \\
    &G(G_0^{-1}-G^{-1})G_0 = GVG_0,
\end{align}
which results in
\begin{align}
    &\frac{1}{E_i-H+i\epsilon} - \frac{1}{E_i-H_0+i\epsilon} =\nonumber \\
    &\frac{1}{E_i-H+i\epsilon}V\frac{1}{E_i-H_0+i\epsilon}.
\end{align}
\bibliographystyle{unsrt}

\end{document}